\documentclass[12pt]{article}
\usepackage[T1]{fontenc}
\usepackage[bbgreekl]{mathbbol}
\usepackage{mathtools}
\usepackage{amssymb,amsmath,mathbbol,mathrsfs}
\DeclareSymbolFontAlphabet{\mathbbm}{bbold}
\DeclareSymbolFontAlphabet{\mathbb}{AMSb}%
\usepackage[mathscr]{euscript}
\usepackage{bm}
\usepackage{xcolor}
\usepackage{tcolorbox}
\usepackage{enumitem}
\usepackage{dsfont}
\usepackage{slashed}
\usepackage{fullpage}
\usepackage{cite}
\usepackage{currvita}
\usepackage[colorlinks=true,linkcolor=blue,citecolor=magenta,linktocpage=true]{hyperref}
\usepackage{braket}
\usepackage{titlesec}
\titleformat*{\section}{\large\bfseries}
\titleformat*{\subsection}{\normalsize\bfseries}
\titleformat*{\subsubsection}{\normalsize\bfseries}
\makeatletter
\renewcommand{\@dotsep}{1000}
\newcommand{\be}{\begin{equation}}
\newcommand{\ee}{\end{equation}}
\newcommand{\bea}{\begin{eqnarray}}
\newcommand{\eea}{\end{eqnarray}}

\newcommand{\Neq}{\stackrel{\sss N}{=}}

\newcommand{\lb}{\label}
\renewcommand{\bar}{\overline}

\newcommand{\cL}{{\mathcal{L}}}

\newcommand{\sE}{\mathscr{E}}
\newcommand{\sJ}{\mathscr{J}}
\newcommand{\sP}{\mathscr{P}}

\newcommand{\sW}{\mathscr{W}}

\newcommand{\sD}{\mathscr{D}}
\newcommand{\sN}{\mathscr{T}}

\renewcommand{\a}{{\alpha}}

\newcommand{\s}{\sigma}

\newcommand{\btheta}{\bar{\theta}}

\newcommand{\mr}{\mathring}
\newcommand{\rd}{\mathrm{d}}
\newcommand{\e}{\mathrm{e}}

\newcommand{\sss}{\scriptscriptstyle}
\newcommand{\mrq}{\mathring{q}}
\newcommand{\mrD}{\mathring{\mathscr{D}}}

\newcommand{\pa}{\partial}

\newcommand{\two}{{}^{\sss (2)}}

\newcommand{\wh}[1]{\widehat{#1}{}}

\begin{document}

\title{\Large{\textbf{\sffamily Carrollian hydrodynamics and symplectic structure on stretched horizons}}}
\author{\sffamily Laurent Freidel$^1$ \& Puttarak Jai-akson$^2$}
\date{\small{\textit{
$^1$Perimeter Institute for Theoretical Physics,\\ 31 Caroline Street North, Waterloo, Ontario, Canada N2L 2Y5\\
$^2$RIKEN iTHEMS, Wako, Saitama 351-0198, Japan\\}}}

\maketitle

\begin{abstract}
The membrane paradigm displays underlying connections between a timelike stretched horizon and a null boundary (such as a black hole horizon) and bridges the gravitational dynamics of the horizon with fluid dynamics. In this work, we revisit the membrane viewpoint of a finite distance null boundary and present a unified geometrical treatment to the stretched horizon and the null boundary based on the rigging technique of hypersurfaces. This allows us to provide a unified geometrical description of null and timelike hypersurfaces, which resolves the singularity of the null limit appearing in the conventional stretched horizon description. We also extend the Carrollian fluid picture and the geometrical Carrollian description of the null horizon,  which have been recently argued to be the correct fluid picture of the null boundary, to the stretched horizon. To this end, we draw a dictionary between gravitational degrees of freedom on the stretched horizon and the Carrollian fluid quantities and show that Einstein's equations projected onto the horizon are the Carrollian hydrodynamic conservation laws. Lastly, we report that the gravitational pre-symplectic potential of the stretched horizon can be expressed in terms of conjugate variables of Carrollian fluids and also derive the Carrollian conservation laws and the corresponding Noether charges from symmetries.
\end{abstract}

\thispagestyle{empty}
\newpage
\setcounter{page}{1}

\hrule
\tableofcontents
\vspace{0.7cm}
\hrule


\section{Introduction}

Boundaries, as hypersurfaces embedded in spacetimes at either finite distances or asymptotic infinities, have been given, in gravitational physics, a special status in present-day theoretical physics. They are no longer treated merely as the loci where boundary conditions are assigned but are now perceived as the locations that birth abundant new and fascinating physics, with the prime examples being the spectacular ideas of gauge/gravity duality in asymptotically anti-de Sitter (AdS) spacetimes \cite{Maldacena:1997re, Witten:1998qj} and celestial holography (see the lecture notes \cite{Strominger:2017zoo,Raclariu:2021zjz,Pasterski:2021rjz} for reviews and references therein) governing infrared physics in asymptotically flat spacetimes. At finite distances, the extensive studies of local subsystems of gauge theories and gravity have unravelled emergent degrees of freedom (usually referred to as edge modes) that encode new (corner) symmetries at the boundaries\cite{Donnelly:2016auv,Speranza:2017gxd,Geiller:2017xad, Geiller:2017whh, Freidel:2020xyx, Freidel:2020svx, Freidel:2020ayo,Ciambelli:2021vnn} and in turn providing a quasi-local holography program for quantizing gravity \cite{Donnelly:2020xgu}.
 This perspective allows for the study of boundary dynamics as generalized conservation laws \cite{Balachandran:1994up,Geiller:2019bti,Freidel:2021cjp} for the corner symmetries charges. However, in this endeavor to unveil the fundamental nature of gauge theories and gravity, different types of boundaries, either null or timelike, have been studied individually depending on the problems at hand and the attempts to seek a unified treatment for them have been scarce. See \cite{Compere:2019bua,  Fernandez-Alvarez:2021yog,Geiller:2022vto} for earlier attempts of unified treatments at infinity.

There exists nonetheless a framework that displays a deep connection between timelike and null surfaces. It is the black hole \emph{membrane paradigm} originated by Damour \cite{Damour:1978cg} and subsequently explored by Throne, Price, and Macdonald \cite{Thorne:1986iy, Price:1986yy}, modeling effectively the physics of black holes seen from outside observers as membranes located at vanishingly close distances to the black hole horizon. These fictitious timelike membranes, which is usually called \textit{stretched horizons}, can also be viewed as arising from quantum fluctuations of geometry around the true horizon (null surface) of the black hole and are furnished with physical quantities such as energy, pressure, heat flux, and viscosity\footnote{The stretched horizon can also be assigned electromechanical properties such as conductance. In this circumstance, one needs to supplement the hydrodynamic equations with some electromechanical equations, such as Ohm's law.}. The intriguing hallmark of the membrane paradigm is that gravitational dynamics of the stretched horizon can be fully written as the familiar equations of hydrodynamics, which in turn allowing us to draw a dictionary between gravitational degrees of freedom and fluid quantities. This profound correspondence, while starting off as a tentative analogy, is a clear reflection of a true nature of gravity and offers a completely hydrodynamic route to gravitational dynamics and opening unprecedented windows to explore some open questions in both fields. Let us also mention that many of its interesting aspects and applications have still been explored in many different contexts, see for example \cite{Anninos:1993zj,Anninos:1994gp,Faulkner:2010jy,Bredberg:2011jq,Bhattacharyya:2015dva}. The fluid/gravity correspondence has been put forth beyond black hole physics in the context of AdS/CFT duality \cite{Bhattacharyya:2007vjd} (see \cite{Son:2007vk,Rangamani:2009xk, Hubeny:2010wp,Hubeny:2011hd} for comprehensive reviews on this topic) and it has been since then generalized and applied in numerous works \cite{Iqbal:2008by,Eling:2009pb,Nickel:2010pr}. It is also worth mentioning other works that uncovered the link between gravitational physics and fluids. Black holes, in many circumstances, actually exhibit droplet-like behaviors akin to liquid. For instance, the Gregory-Laflamme instability of higher-dimensional black strings \cite{Gregory:1993vy} displays similar behavior to the Rayleigh instability of liquid droplets \cite{Cardoso:2006ks}. The work \cite{Freidel:2014qya} also showed that dynamics of a timelike surface (which they called gravitational screen) behaves like a viscous bubble with a surface tension and an internal energy. Analog models of black holes \cite{Unruh:1980cg} illustrated the converse notion and argued that kinematic aspects of black holes can be reproduced in hydrodynamical systems and that fluids can admit sonic horizons and even the analog of Hawking temperature. Lastly, in the context of local holography, the corner symmetry group of gravity was shown to contain the symmetry group of perfect fluids as its subgroup \cite{Donnelly:2020xgu}. Furthermore, the advantage of treating timelike surfaces and null surfaces in the same regard stems from the observation that some information of null boundaries, which are true physical boundaries are seemingly obtained when considering small deviations from those boundaries. In other words, those information can only be accessed by considering timelike surfaces located near the boundaries. This lesson has been demonstrated explicitly at asymptotic null infinity at which the radial ($1/r$) expansion around null infinities encodes higher-spin symmetries and conservation laws of the null infinities \cite{Freidel:2021qpz,Freidel:2021dfs,Freidel:2021ytz}. 

One issue of the stretched horizon description of a null boundary is that the horizon energy-momentum tensor and its conservation laws, which require a notion of induced metric and connection, on the stretched horizon are singular when evaluated on the null boundary due to the infinite redshift.
In the original membrane paradigm perspective, the singularities of the horizon fluids are  remedied by considering  an ad-hoc renormalized (red-shifted) version of those quantities \cite{Damour:1978cg,Thorne:1986iy, Price:1986yy}. 
This  \emph{null limit} from the stretched horizon to the null boundary was recently argued by Donnay and Marteau \cite{Donnay:2019jiz} to coincide with the Carrollian limit à la Lévy-Leblond \cite{Leblond1965} and that the corresponding membrane fluids are Carrollian \cite{Ciambelli:2018xat,Ciambelli:2018ojf,Petkou:2022bmz,Freidel:2022bai}, rather than  relativistic or non-relativistic fluids (see also \cite{Penna:2018gfx} for an early argument). 

This non-smooth null limit obstructs us from uncovering a precise connection between the hydrodynamic and geometrical picture of the timelike stretched horizon and the null boundary. Also, the link between various constructions in the null case and the timelike case have never been fully made precise. This means that  conclusions we  reached in the null case can not be made in the timelike case, and vice versa. This especially includes the disparity in the construction of the energy-momentum tensor and its conservation laws.
In the timelike case the energy momentum tensor and gravitational charges of the surfaces can be constructed using the Brown-York prescription \cite{Brown:1992br, Brown:2000dz}. Moreover the conservation laws are usually written in terms of the Levi-Civita connection on the hypersurface.

The null case is on the other hand 
more subtle. One important subtlety is that there is no notion of Levi-Civita connection on a null surface. Another one is that the usual definition of a strong Carrollian connection used in \cite{Bekaert:2015xua,Figueroa-OFarrill:2019sex,Figueroa-OFarrill:2020gpr, Herfray:2021qmp, Ashtekar:2021wld,Baiguera:2022lsw, Hansen:2021fxi}, which works well for asymptotic null infinity, is too restrictive to deal with finite distance null surfaces. 
As a result, a lot of efforts have been put into the understanding of the phase space, the notion of energy-momentum tensor, and conserved charges of the null surfaces \cite{Donnay:2015abr,Donnay:2016ejv,Chandrasekaran:2018aop,Hopfmuller:2018fni, Jafari:2019bpw, Chandrasekaran:2020wwn,Adami:2021nnf,Adami:2021kvx,Chandrasekaran:2021hxc}.
In addition,  there exist ample evidences suggesting a correspondence between the geometry and physics at null boundaries and Carrollian theories, both in finite regions \cite{Ciambelli:2019lap,Chandrasekaran:2021hxc} and at infinities \cite{Duval:2014uva,Duval:2014lpa,Bagchi:2019xfx,Ciambelli:2019lap,Herfray:2021qmp,Bagchi:2022emh,Campoleoni:2022wmf,Bagchi:2022owq, Donnay:2022aba, Donnay:2022sdg,Gupta:2020dtl,Bagchi:2022eav}.
What is missing is a unified geometrical treatment of null and timelike stretched horizon. One difficulty is that the connection used in the conservation laws of the hypersurface energy-momentum tensor is radically different in the timelike and null cases. Resolving these issues by seeking for a unified treatment of these two types of hypersurfaces (or boundaries) that admits a smooth null limit is the main goal of this work. 

The objectives, the outline, and some key results of this article are presented below. 
\begin{enumerate}[label = \roman*)]

\item \emph{Removal of the singularity of the membrane paradigm:} As we have already mentioned, the main issue hindering the link between various geometrical constructions and the fluid picture presented at the stretched horizon and the true null horizon is the presence of the singular limit in the standard Brown-York formalism for timelike surfaces. To cure this, we extend the construction of Chandrasekaran  et al. \cite{Chandrasekaran:2021hxc} and
utilize the rigging technique \cite{Mars:1993mj, Mars:2013qaa} to construct a hypersurface connection on stretched horizons which admits a non-singular limit to the null boundary. 
 We show in section \ref{sec-rigged} that the geometry of the stretched horizon descending from this technique admits a non-singular limit to the null boundary, therefore providing a unified description for both timelike and null hypersurfaces. We then construct the energy-momentum tensor $T_a{}^b$, from the geometrical data of the surfaces and show that its conservation laws are the Einstein's equations projected onto the stretched horizon, 
\begin{align}
D_b T_a{}^b = \Pi_a{}^b G_{b}{}^c n_c\hat{=}0,
\end{align}
where $n_a$, $\Pi_a{}^b$, and $D_a$ are respectively the normal to the stretched horizon, the rigged projector, and the rigged connection on the horizon. All of them are regular on the null boundary, consequently providing a non-singular stretched horizon viewpoint to the null boundary. Our construction hence generalizes the previous results for the null case \cite{Parikh:1997ma,Hopfmuller:2018fni,Jafari:2019bpw,Chandrasekaran:2021hxc,Adami:2021nnf}. Precise definitions and details are provided in section \ref{sec:cons}.

\item \emph{Carroll structures and Carrollian hydrodynamics on timelike surfaces:} While it has been established that Carroll geometries are natural intrinsic geometries of null surfaces, both at finite and infinite regions \cite{Chandrasekaran:2018aop, Chandrasekaran:2021hxc, Ashtekar:2021wld}, it has never been known how to assign the notion of Carrollian to the geometry of timelike surfaces. One of the key idea we would like to convey in this work is that the rigged structure endowed on the stretched horizon naturally induces a geometrical Carroll structure on the stretched horizon. 
It is important to appreciate that by a geometrical Carroll structure on a stretched horizon we follow the definition of Ciambelli et al. \cite{Ciambelli:2019lap}: 
By a \emph{geometrical Carroll structure} we mean the existence of a line bundle over a 2-sphere equipped with a metric. The vertical lines of the bundle define a congruence of curves tangent to the Carrollian vector $\ell$. The pull-back of the 2-sphere metric defines a null metric $q$ on the 3-dimensional manifold. This metric can differ from the stretched horizon induced metric by at most a rank one tensor.
The notion of a geometrical Carroll structure is central to the description of fluids in the Carrollian limit, see \cite{Ciambelli:2018xat,Petkou:2022bmz,Freidel:2022bai}.

This notion of a geometrical Carroll structure is weaker than the usual notion of a \emph{strong Carroll structure} or what we refer to as a Carroll G-structure. A \emph{Carroll G-structure} consists of a geometrical Carroll structure together with a connection compatible with the bundle structure and the base metric. The defining condition for this connection is that its  structure group is the Carroll group. Such a connection is called a strong Carrollian connection.
This is the notion used in \cite{Bekaert:2015xua,Figueroa-OFarrill:2019sex,Figueroa-OFarrill:2020gpr, Herfray:2021qmp, Ashtekar:2021wld,Baiguera:2022lsw, Hansen:2021fxi}. The notion of Carroll G-structure is too strong for the description of stretched horizon. However, stretched Horizons can be equipped with a geometrical Carroll structure and a torsionless connection which only preserves the base metric even if they are not null. 

Interestingly, the difference between a non-null stretched horizon and its null limit can be seen in the structure of its energy-momentum tensor $T_a{}^b$. The Carrollian fluid energy current is  given by  $-\ell^a T_a{}^b =\sE \ell^b + \sJ^b$, where $\sE$ is the fluid energy density and $\sJ^b$ is the heat flow current tangent to the surface. 
It turns out that when the stretched horizon is null, the heat flow has to vanish while for a general stretched horizon, the heat current is simply proportional to the fluid momenta. As we will see, these relations are simply the expression of the boost symmetry which differs on null and timelike surfaces \cite{Baiguera:2022lsw}.
 We will also show in section \ref{sec:cons} that the Einstein equations on the stretched horizon can be written exactly as the evolution equations of the energy density and momentum density of Carrollian hydrodynamics. 

\item \emph{Gravitational phase space is Carrollian:} Lastly, in section \ref{sec:symplectic}, we will evaluate the pre-symplectic potential, capturing the information of the gravitational covariant phase space, on the stretched horizon and show that it can be expressed in terms of the conjugate variables of Carrollian fluids \cite{Freidel:2022bai}, 
\begin{align}
\Theta^{\text{can}}_H [g, \delta g]  = \delta S_{\text{fluid}} - \int_H \btheta \delta \rho. 
\end{align} 
Here $S_{\text{fluid}}$ is the Carrollian fluid action whose variation under the stretched horizon geometrical structure defines the energy-momentum tensor. The stretched horizon contains an extra term in addition to the null horizons \cite{Parattu:2015gga,Parattu:2016trq,Hopfmuller:2016scf}: $\rho$ is a scalar that measures the non-nullness of the stretched horizon and $\bar\theta$ is its transverse expansion.

\end{enumerate}

\noindent \textbf{Notations and conventions:} In this work, we adopt the gravitational unit where $8\pi G =1$. The notations we will use are listed below.
\begin{itemize} [topsep=0pt,itemsep=0pt,partopsep=0pt, parsep=0pt]
\item Small letters $a,b,c,...$ are spacetime indices. They are raised and lowered by the spacetime metric $g_{ab}$ and its inverse $g^{ab}$.
\item Capital letters $A,B,C,...$ are horizontal (or sphere) indices. They are raised and lowered by the 2-sphere metric $q_{AB}$ and its inverse $q^{AB}$.
\item Spacetime differential forms are denoted with boldface letters such as $\bm{k}, \bm{n}, \bm{\omega}, \bm{\epsilon},...$
\item The wedge product between differential forms is denoted by $\wedge$ as usual while $\odot$ is used to denote symmetric tensor product, that is $A \odot B = \frac{1}{2}(A \otimes B + B \otimes A)$.
\item Directional derivative of a function $f$ along a vector field $V$ is written as $V[f] = V^a \pa_a f$. 
\item We sometimes adopt index-free notations. For example, the inner product between a vector $X$ and a vector $Y$ computed with the metric $g$ is written as $g(X,Y) = g_{ab}X^a Y^b$. 

\end{itemize}


\section{Geometries of stretched horizons and null boundaries} \lb{sec-rigged}

We dedicate this section to lay down relevant geometrical constructions of null and timelike hypersurfaces, focusing particularly on the case of finite distance surfaces. The physical examples of them are event horizons of black holes (null boundaries) and fictitious stretched horizons (timelike surfaces) located at small distances outside the black hole horizons.

Geometrical construction of hypersurfaces usually depends on the type of hypersurfaces and problems under consideration. For instance, the Arnowitt-Deser-Misner (ADM) formalism, centering around the (3+1)-decomposition of spacetime, has become a go-to tool to deal with spacelike Cauchy surfaces and timelike boundaries. This (3+1)-splitting approach relies on the existence of the apparent notion of time (through the spacelike foliations of spacetime) and is thus useful when one wants to tackle initial-value problems of general relativity or study Hamiltonian formulation of general relativity (see for instance \cite{Gourgoulhon:2007ue} and references therein). The analog of this formalism for null hypersurfaces has been considered in \cite{Gourgoulhon:2005ng}. This ``time-first'' formalism instinctively imprints Galilean nature to the considerations, rather than the Carrollian nature which is a ``space-first'' constructions. In this regards, we thus refrain from adopting the ADM formalism in our study. In the case of a null hypersurface, the spacetime geometry in close vicinity to the surface has been studied extensively using the Gaussian null formalism which utilizes null geodesics to extend the intrinsic coordinates on the null surface to the surrounding spacetime and it has been used to describe the near-horizon geometry of black holes \cite{Donnay:2015abr, Donnay:2016ejv, Donnay:2019jiz} and also geometry of general null surfaces located at finite distances
\cite{Hopfmuller:2016scf,Hopfmuller:2018fni,Adami:2021nnf,Adami:2021kvx}. Another type of framework suitable for studying the geometry of null hypersurfaces is the double null foliation technique \cite{Brady:1995na}, which is a spacial (gauge fixed) case of a more general formalism, the (2+2)-splitting formalism. The (2+2)-splitting of spacetime has been proven to be the most apt formalism for describing the geometry around codimension-2 corner spheres, regardless of the nature of codimension-1 boundaries, and has been tremendously utilized in the arena of local holography program \cite{Donnelly:2016auv,Hopfmuller:2016scf,Donnelly:2020xgu}. In the context of asymptotic null infinity, the Bondi-Metzner-Sachs (BMS) formalism, the Bondi gauge and its extensions \cite{Bondi:1962px, Sachs:1962wk, Freidel:2021fxf, Freidel:2021qpz, Freidel:2021dfs} as well as the Newman-Unti gauge \cite{Newman:1962cia} (see also \cite{Geiller:2022vto} for the enlarged gauge choice) have been widely adopted. Intrinsically, the geometry of null surfaces can also be understood from the perspective of Carroll geometries \cite{Duval:2014uoa,Duval:2014uva,Duval:2014lpa,Ciambelli:2019lap}. Here, we seek for the kind of general geometrical construction that works for all types of hypersurfaces. To this end, we will adopt the rigging technique \cite{Duggal,Mars:1993mj,Mars:2013qaa} and will show that it delivers a unified geometrical construction that treats timelike and null surfaces on an equal footing, which admits a smooth null limit.

To set a stage, we consider a region of a 4-dimensional spacetime $M$, endowed with a Lorentzian metric $g_{ab}$ and a Levi-Civita connection $\nabla_a$, that is bounded by a null boundary $N$ located at a finite distance. It is then foliated into a family of 3-dimensional timelike hypersurfaces, \emph{stretched horizons} $H$, situated at constant values of a foliation function $r(x) = \text{constant} >0$. Situated at $r(x) =0$ is the null boundary $N$. In this setup, the \emph{null limit} from the stretched horizon $H$ to  the null boundary $N$ simply corresponds to the limit $r \to 0$. 

In practice, another foliation function is introduced to further provide a time slicing structure to the spacetime $M$, and together with the radial function $r(x)$ establishes the $(2+2)$-decomposition of spacetime \cite{Donnelly:2016auv,Hopfmuller:2016scf, Hopfmuller:2018fni,Donnelly:2020xgu}, in turn rendering a notion of time apparent. Doing so would inevitably bring to the surfaces $H$ and $N$ the Galilean picture. However, we will not adopt this technique. Instead, we seek for the Carrollian viewpoint by considering the surface $H$ (and also the boundary $N$ as a limit) as a fiber bundle, $p:H \to S$, where the space $S$ is chosen to be a 2-sphere with local coordinates $\{ \s^A \}$ and a sphere metric $q_{AB} \bm{\rd} \s^A \odot \bm{\rd} \s^B$. The surface $H$ then admits a Carroll structure \cite{Duval:2014uoa,Ciambelli:2019lap,Freidel:2022bai}. \\

\noindent\fbox{%
\parbox{\textwidth}{%
\textbf{Carroll structures:}
A (weak) Carroll structure is given by a triplet $(H, \ell, q)$ where a vector field $\ell$, called the Carrollian vector field, points along a fiber, meaning that $\ell \in \text{ker}(\bm{\rd} p)$, and a null Carrollian metric $q$ is a pullback of the sphere metric, $q = p^* (q_{AB} \bm{\rd} \s^A \odot \bm{\rd} \s^B)$ satisfying the condition $q(\ell, \cdot) =0$. 
}} \\

While the stretched horizon $H$ does not has the temporal-spatial split, its tangent space $TH$ does admit, as inherited from the fiber bundle structure, the vertical-horizontal split, which is determined by an Ehresmann connection 1-form $\bm{k}$ dual to the Carrollian vector $\ell$, i.e., $\iota_\ell \bm{k} =1$. The Ehresmann connection allows us to select a horizontal distribution whose basis vectors are denoted $e_A$ and satisfy $\iota_{e_A} k=0$. We will elaborate more about Carroll structures later. Let us mention here that the structure constants of the Carroll structure are given by the acceleration $\varphi_A$ and the vorticity $w_{AB}$ which enters the vector fields commutation relations 
\begin{align}
[\ell, e_A] = \varphi_A \ell, \qquad [e_A, e_B] = w_{AB} \ell.  \lb{com-Carr}
\end{align}

The key concept we would like to demonstrate in this section is that a Carroll structure is a natural intrinsic structure of the stretched horizon $H$, and is inherited from a rigged structure, a type of extrinsic structure to $H$ which we will discuss shortly, and together, they fully describe the complete geometry of $H$. Let us highlight again here that our construction holds for both timelike and null hypersurfaces and the null limit is non-singular, which therefore provides a unified treatment of these hypersurfaces. Let us finally describe in detail our geometric construction of the stretched horizon.

\subsection{Rigged Structures}
\begin{figure}[t]
\centering
\includegraphics[scale=0.3]{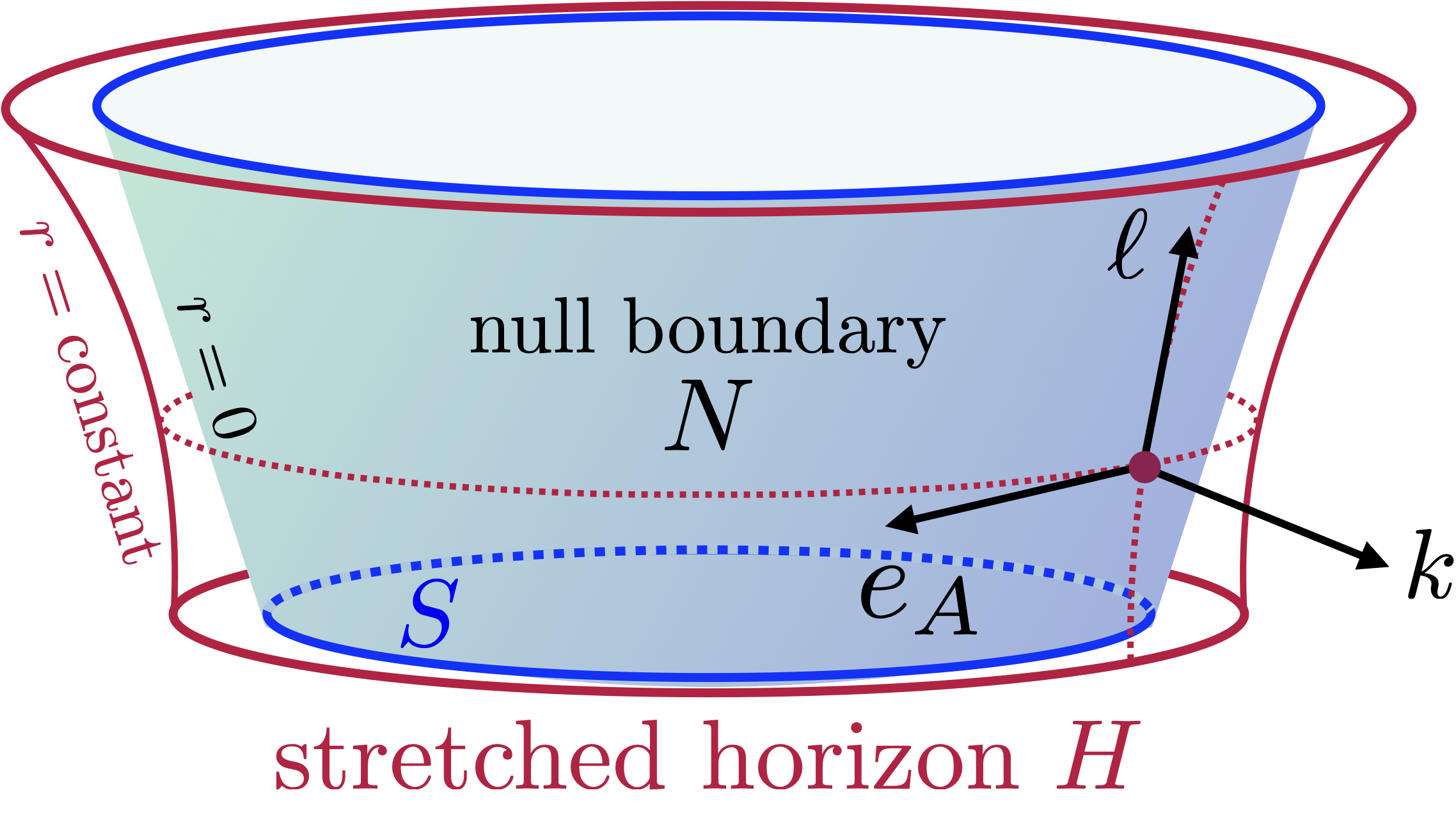} 
\caption{Stretched horizons $H$ are chosen to be hypersurfaces at $r = \text{constant}$ and the null boundary $N$ is the limit $r \to 0$ of the sequence of stretched horizons. The surface $H$ is endowed with the rigging vector $k$ and its dual form $\bm{n}$. The Carroll structure with the vertical vector $\ell$ and the horizontal vector $e_A$ is induced from the rigged structure, and together with $k$, they form a complete basis for the tangent space $TM$.} \lb{stretched}
\end{figure}
 
We begin with the introduction of a rigged structure  \cite{Duggal,Mars:1993mj,Mars:2013qaa}, which provides an extrinsic structure of the stretched horizon $H$. Recalling that $H$ is embedded in the spacetime at the location $r=\text{constant}$, it is then equipped with a normal form $\bm{n}=n_a \bm{\rd} x^a$.
This means any vector field $X$ tangent to the surface $H$ is such that $\iota_X \bm{n}=0$. 
We consider the normal form that defines a foliation of the ambient spacetime $M$, meaning that $\bm{\rd n} = \bm{a} \wedge \bm{n}$ for a 1-form $\bm{a}$ on $M$. In this setup, the normal form is given by
\begin{align}
\bm{n} = \e^{\bar{\alpha}} \bm{\rd} r, \lb{n-def}
\end{align}
for a function $\bar{\alpha}$ on $M$ and correspondingly we have that $\bm{a} = \bm{\rd} \bar{\alpha}$ as desired. 

To describe the geometry of the stretched horizon, we adopt the rigging technique of a general hypersurface \cite{Mars:1993mj,Mars:2013qaa} and endow on $H$ a \emph{rigged structure} given by a pair $(\bm{n}, k)$, where $\bm{n} $ is the aforementioned normal form and a \emph{rigging vector} $k = k^a \pa_a$ is transverse to $H$ and is dual to the normal form,
\begin{align}
\iota_k \bm{n} = 1.
\end{align}
With this, we next define the rigged projection tensor, $\Pi: TM \to TH$, whose components are given in terms of the rigged structure by
\begin{align}
\Pi_a{}^b := \delta_a^b - n_a k^b, \qquad \text{such that} \qquad k^a \Pi_a{}^b = 0 = \Pi_a{}^b n_b. \lb{projector}
\end{align}
This rigged projector is designed in a way that, for a given vector field $X$ on $M$, the vector $\bar{X}^b := X^a \Pi_a{}^b \in TH$ is tangent to $H$ with $\bar{X}^a n_a =0$. 
Similarly, for a given 1-form $\bm{\omega} \in T^*M$, the 1-form $\bar{\omega}_a := \Pi_a{}^b \omega_b \in T^*H$ is such that $k^a \bar{\omega}_a=0$.

\subsection{Null rigged structures and Induced Carroll Structures}

Equipping the spacetime $M$ with a Lorentzian metric $g = g_{ab} \bm{\rd} x^a \odot \bm{\rd} x^b$ and its inverse $g^{-1} = g^{ab}\pa_a \odot \pa_b$ let us define the 1-form $\bm{k} = g(k, \cdot)$ and the vector $n = g^{-1}(\bm{n}, \cdot)$. We can also define the transverse 1-form $\bar{k}_a = \Pi_a{}^b k_b$ such that $\bar{k}_a k^a =0$. There exists different type of rigged structures depending on the nature of the rigging vector $k$. For timelike surfaces, one usually adopts the choice where $\bar{k}_a=0$. This choice corresponds to a normal rigged structure such that $ k_a = n_a/ |n|^2$ where the norm $|n|^2 := n_a n^a \Neq 0$ vanishes on the null boundary. This rigged structure is obviously singular when the surface is null and is the source of all singularities encountered when considering the null limit of the induced connection and the induced energy-momentum tensor in the membrane paradigm framework. Another choice, which we will adopt in this work and is regular for both timelike and null cases, is a \emph{null rigged structure}. It is the case where $\bar{k}_a =k_a$ which also infers that the rigging vector $k$ is null. Denoting by $2\rho$ the norm-square of the normal 1-form, we overall have the following conditions,
\begin{align}
g(k,k)=0,\qquad g^{-1}(\bm{n}, \bm{n}) = n_a n^a := 2\rho.
\end{align} 
It is always possible to adjust the factor $\bar{\alpha}$ defined in \eqref{n-def} to insure that the norm $\rho$ stays constant on the stretched horizons $H$, i.e., $\Pi_a{}^b \pa_b \rho =0$. 
 As we will see later, this is going to be important for the construction of the surface energy-momentum tensor. 
 
We define a tangential vector field $\ell = \ell^a \pa_a \in TH$ whose components are given by the projection of the vector $n^a$ onto the surface $H$, i.e., $\ell^a := n^b \Pi_b{}^a $. Using the definition of the projector \eqref{projector}, one can check that this tangential vector is related to the vector $n$ and $k$ by 
\begin{align}\label{defnl}
n^a = 2\rho k^a + \ell^a. 
\end{align}
Furthermore, one can easily verify that the vector $\ell$ and the 1-form $\bm{k}$ obey the following properties,
\begin{align}
\qquad \iota_\ell \bm{n} =0, \qquad \text{and} \qquad \iota_\ell \bm{k} = 1. 
\end{align}

While the first property stems from the definition that $\ell$ is tangent to the surface $H$, the second property $\iota_\ell \bm{k} =1$ readily suggests that we can treat the tangential vector $\ell$ as an element of a Carroll structure on $H$, and the 1-form $\bm{k}$ plays a role of a Ehresmann connection that defines the vertical-horizontal decomposition of the tangent space $TH$ (see the detailed explanation in \cite{Ciambelli:2019lap,Freidel:2022bai}). Other objects that belong to the Carroll geometry, including the horizontal basis $e_A$ and the co-frame field $\bm{e}^A$, follow naturally from this construction. To see this, one uses that the projector can be further decomposed as 
\begin{align}
\Pi_a{}^b = q_a{}^b + k_a \ell^b, \qquad  \text{with} \quad q_a{}^b k_b =0=\ell^a q_a{}^b.
\end{align}
The tensor $q_a{}^b = e^A{}_a e_A{}^b$ is the horizontal projector from the tangent space $TH$ to its horizontal subspace. The last element of the Carroll structure, the null Carrollian metric on $H$, is given by $q_{ab} =q_a{}^c q_b{}^d g_{cd}$. We will also make an additional assumption that the projection map, $p: H \to S$, stays the same for all $H$, inferring that the co-frame $\bm{e}^A$ on $H$ is closed, $\bm{\rd}\bm{e}^A =0$, throughout the spacetime $M$. 



It is important to appreciate the result we have just developed --- a Carroll structure on the space $H$ is fully determined from the rigged structure and the spacetime metric. Let us summarize again all important bits in the box below (see Appendix \ref{App:Carr} for relevant details).\\

\noindent\fbox{%
\parbox{\textwidth}{%
\textbf{\underline{Induced Carroll structure}:} Given a \emph{null} rigged structure $(k, \bm{n})$ on a hypersurface $H$, with the rigged vector field $k$ being null, and the spacetime metric $g$, the Carroll structure $(H, \ell, q)$ is naturally induced on the hypersurface. The vertical vector field $\ell$ and the Ehresmann connection $\bm{k}$ are related to the rigged structure by
\begin{equation}
\begin{aligned}
\ell^a =  n_c g^{cb}\Pi_b{}^a, \qquad \text{and} \qquad k_a = g_{ab} k^b. 
\end{aligned}
\end{equation}
 The null Carrollian metric is $q_{ab} = q_a{}^c q_b{}^d g_{cd}$, where $q_a{}^b = \Pi_a{}^b - k_a \ell^b$ is a horizontal projector.
 }%
}
\\

The vectors $(\ell, k ,e_A)$ and their dual 1-forms $(\bm{k}, \bm{n}, \bm{e}^A)$ thus span the tangent space $TM$ and the cotangent space $T^*M$, respectively (see Figure \ref{stretched}). The ambient spacetime metric decomposes in this basis as 
\begin{equation}
\lb{4metric}
\begin{aligned}
g_{ab} &=  q_{ab} + k_a \ell_b + n_a k_b   \\
&=  q_{ab} + 2 n_{(a} k_{b)}   - 2\rho k_a k_b. 
\end{aligned}
\end{equation}
Let us also observe that, in general, the Carrollian vector field $\ell$ is not null and its norm is 
\begin{align}
\ell_a\ell^a=-2\rho. 
\end{align}
This expresses the fact that the Carroll structure is null strictly on the null boundary $N$. Note that the metric expression is regular when $\rho=0$, and we have on the null boundary that $n_a \Neq \ell_a$.

Armed with the induced Carroll structure on $H$, almost all analysis done in the previous literature can be applied. One however has to keep in mind that rather than considering the space $H$ only on its own, viewing $H$ as a surface embedded in the higher-dimensional spacetime equips us with richer geometry. In our consideration, this additional geometry arises from the transverse direction, capturing by the rigged structure $(k, \bm{n})$. 


To simplify our computations, let us make another assumption that the null transverse vector $k$ generates null geodesics on the spacetime $M$, meaning that $\nabla_k k = \bar\kappa k$.\footnote{We do not impose that the geodesic is affinely parameterized because we want to keep the rescaling symmetry $(\ell,k)\to (\Omega \ell,\Omega^{-1} k)$ alive. Under this symmetry we have that 
\be
\bar\kappa \to \Omega^{-1}(\bar\kappa - k[\Omega]).
\ee 
 Using the rescaling symmetry we can always achieve that $\bar\kappa=0$ which restricts the rescaling symmetry to be such that $k[\Omega]=0.$  
 }
 This particularly infers that the curvature of the Ehresmann connection does not contain the normal direction\footnote{This is also equivalent to the condition $\iota_k \bm{\rd k} = \cL_k \bm{k} = \bar\kappa \bm k$, and one can check, following from the null-ness property of $k$, that $k^a (\bm{\rd} k)_{ab} = \nabla_k k_a$.},  
\be
\bm{\rd k} := \bar\kappa \bm{n} \wedge \bm{k} - \varphi_A (\bm{k} \wedge \bm{e}^A) - \frac{1}{2}w_{AB} (\bm{e}^A \wedge \bm{e}^B), \lb{d-k-4}
\ee 
where the components $\varphi_A$ and $w_{AB}$ are Carrollian acceleration and the Carrollian vorticity, respectively. Let us also recall that we have chosen earlier the null normal $\bm{n} = \e^{\bar{\alpha}} \bm{\rd} r$ to defines a foliation of the spacetime $M$. The curvature of the normal form is
\begin{align}
\bm{\rd n} = \ell[\bar{\alpha}]\bm{k} \wedge \bm{n} -e_A[\bar{\alpha}] \bm{n} \wedge \bm{e}^A, \lb{d-n}
\end{align}
The components $\ell[\bar{\alpha}]$ and $e_A[\bar{\alpha}]$, as we will see momentarily, are related to the surface gravity and the Hájíček 1-form field of the surface. Let us also mention again that the curvature $\bm{\rd e}^A  =0$ by construction. 

The curvatures of the basis 1-forms determine the commutators of their dual vector fields\footnote{The relation is $\iota_X \iota_Y \bm{\rd \omega} = \iota_{[X,Y]}\bm{\omega} + \cL_Y (\iota_X \bm{\omega}) -\cL_X (\iota_Y \bm{\omega})$ for a 1-form $\bm{\omega}$ and vector $X$ and $Y$. }. In this case, it follows from \eqref{d-k-4} and \eqref{d-n} that the non-trivial commutators of the basis vector fields are  
\begin{align}
[\ell, e_A] = \varphi_A \ell, \qquad [e_A, e_B] = w_{AB} \ell, \qquad [k,\ell] = \ell[\bar{\alpha}] k - \bar\kappa \ell,\qquad [k, e_A] = e_A[\bar{\alpha}] k.  \lb{com-4}
\end{align}
The first two terms again are the Carrollian commutation relations \eqref{com-4}. 

\subsection{Local Boost and rescaling symmetries}

Let us emphasize that the rigged structure $\Pi_a{}^b$ is invariant under a rescaling symmetry 
\be \label{rescaling}
\ell \to e^\epsilon \ell, \qquad k\to e^{-\epsilon} k, \qquad q_{ab}\to q_{ab}.
\ee 
Under this symmetry we have that 
\be 
\bar\alpha \to \bar\alpha +\epsilon, 
\qquad
\rho \to e^{2\epsilon}\rho,\qquad 
\varphi_A \to \varphi_A -e_A[\epsilon],\qquad
\bar\kappa\to e^{-\epsilon}(\bar\kappa -k[\epsilon]).
\ee
On one hand, the transverse dependence of this  symmetry can be fixed by imposing that the geodesics are affinely parameterized.  On the other hand the tangential dependence of this symmetry can be fixed by demanding that $\rho$ is constant on a given surface $H$. As we will see later, the second condition will play a crucial role when imposed on all stretched horizons.
For the moment, we leave this symmetry unfixed as this provides a nice consistent check on the conservation equations satisfied by the rigged geometry.

Besides the rescaling symmetry, the decomposition of the bulk geometry $g_{ab}$ in terms of the geometry of stretched horizon $(q_{ab}, \ell^a, k_a, n_b)$ possesses another local symmetry, the boost symmetry, that preserves the spacetime metric $g_{ab}$. While the rescaling symmetry preserves the rigged structure the boost symmetry does. The rescaling symmetry labelled by a parameter $\epsilon$ is
simply given by 
\be 
\delta_\epsilon n_a=\epsilon n_a, \quad 
 \delta_\epsilon k_a =-\epsilon k_a,\quad  \delta_\epsilon \ell^a= \epsilon \ell^a, \quad
 \delta_\epsilon q_{ab} = 0.
\ee 
It preserves the rigged structure. 
The boost symmetry is labelled by a vector $\lambda^a$ which is horizontal, meaning that $\lambda^a n_a = \lambda^a k_a=0$.
The infinitesimal boost transformation acts as 
\be 
\delta_\lambda n_a=0,\qquad \delta_\lambda k_a= \lambda_a,
\qquad \delta_\lambda \ell^a =-2\rho \lambda^b,
\quad\delta_\lambda q^{ab} = -(\lambda^a \ell^b + \ell^a \lambda^b).\lb{boost}
\ee 
This transforms the rigged projector as $\delta_\lambda\Pi_a{}^b = - n_a \lambda^b$ while preserving $g_{ab}$. When $\rho=0$ on the null boundary $N$, the boost symmetry leaves the Carrollian vector $\ell$ invariant (see for instance \cite{Baiguera:2022lsw}).

\subsection{Coordinates}

We now supplement our geometrical construction of intrinsic structure of stretched horizons with the introduction of coordinates. As we have set up that the stretched horizons $H$ are defined to be hypersurfaces labelled by a parameter $r \geq 0$, we can chose $r$ to serve as a radial coordinate. Furthermore, let us use $(u,y^A)$ as general coordinates on $H$ and they are chosen so that a cut at constant $u$ is identified with a sphere $S$. The coordinates $(u,y^A)$ are then extended throughout the spacetime $M$ by keeping their values fixed along null geodesics generated by the transverse vector $k$. Overall, we adapt $x^a = (u,r,y^A)$ as the coordinates on the spacetime $M$. 

In this coordinate system, the basis vector fields are expressed as follows (we follow the parameterization for the tangential basis from our precursory work \cite{Freidel:2022bai})
\begin{align}
\ell = \e^{-\a} D_u, \qquad k = \e^{-\bar{\alpha}}\pa_r \qquad e_A = (J^{-1})_A{}^B \pa_B + \beta_B D_u
\end{align}
where we defined $D_u = \pa_u + V^A \pa_A$. The corresponding dual basis 1-forms are given by
\begin{align}
\bm{k} = \e^\a (\bm{\rd} u  - \beta_A \bm{e}^A), \qquad \bm{n}= \e^{\bar{\alpha}} \bm{\rd} r ,\qquad \bm{e}^A = (\bm{\rd }y^B - V^B \bm{\rd} u) J_B{}^A. 
\end{align}

The components $(\beta_A, V^A, J_A{}^B)$ that are parts of the Carroll geometry are functions of the coordinates $(u,y^A)$ on the stretched horizon $H$. We note again that $\bm{e}^A$ is given as the pullback of $\rd \sigma^A$ by the bundle map $p:H\to S$, where $\sigma^A $ are local coordinates on the base space $S$. Their independence of the radial coordinate $r$ stems from our construction that the Carroll projection $p: H \to S$ is independent of the foliation defined by the function $r(x)$, and that $k$ is tangent to null geodesics. One can indeed be more general by relax the $r$-independent conditions. Doing so would inevitably introduce more variables, i.e., radial derivatives of these components, to the consideration which thereby renders the computations more complicated. We refrain from doing so and keep our analysis simple in this article. Let us also remark that, even though the frame $\bm{e}^A$ is set to be independent of the radial direction, the null Carrollian metric $q_{ab}$ can still depend on $r$ due to the possible $r$-dependence of the sphere metric $q_{AB}$. The remaining metric components, which are the norm $\rho$ and the scales $\alpha$ and $\bar{\alpha}$, are in general functions of $(u,r,y^A)$. We will however impose in the following section that $\rho$ only depends on $r$, that is $D_a \rho =0$ for the reason we will justify momentarily.
The metric in coordinates is given by 
\bea
\rd s^2 &=& 2\bm{k}\odot( \e^{\bar{\alpha}} \bm{\rd} r -\rho \bm{k} ) + \tilde{q}_{AB} (\bm{\rd} y^A -V^A \bm{\rd} u) \odot ( \bm{\rd} y^B - V^B \bm{\rd} u),
\eea
where $\tilde{q}_{AB}= J_A{}^CJ_B{}^D q_{CD}$.
It assumes the Bondi form \cite{Bondi:1962px, Sachs:1962wk} if we impose that $\beta_A=0$ which means that $\bm{k}=e^\alpha \rd u$. 
It assumes the Carrollian form \cite{Ciambelli:2019lap} if we choose co-moving coordinates $y^A=\sigma^A$ for which $V^A=0$. Let us note that the induced metric on the stretched horizon takes the Zermelo form when $\beta_A =0$ and it takes the Randers-Papapetrou form when $V^A =0$ \cite{Gibbons:2008zi,Ciambelli:2018xat}.

\subsection{Rigged metric, rigged derivative and rigged connection}\label{sec:con}

Provided the rigged structure on the stretched horizon $H$, we can define the rigged metric, $H_{ab} := \Pi_a{}^c\Pi_b{}^d g_{cd}$, and its dual, $H^{ab} := g^{cd} \Pi_c{}^a\Pi_d{}^b$. Given any two tangential vectors $X,Y\in TH$ that, by definition, satisfy the condition $X^a n_a = Y^a n_a =0$, we can clearly see that
\begin{align}
H_{ab} X^a Y^b =  g_{ab} X^a Y^b,\qquad \text{and} \qquad H_{ba} k^a =0.
\end{align}
This shows that the rigged metric $H_{ab}$ acts on tangential vector fields the same way as the induced metric $h_{ab} = g_{ab} - \frac{1}{2\rho} n_a n_b$. The difference, however, lies in the fact that the induced metric is orthogonal $h_{ab}n^b=0$ while the rigged metric 
satisfy the transversality condition $H_{ab} k^b=0$.
Combining this definition with \eqref{4metric} we see that the rigged  metric  on the space $H$, and its dual, can be written in terms of the Carroll structure as 
\begin{align}
H_{ab} = q_{ab} - 2\rho k_a k_b, \qquad \text{and} \qquad H^{ab} = q^{ab}, 
\end{align}
Observe that the advantage of the rigged metric is that it provides an expression which is regular when taking the null limit, $\rho \to 0$, while, on the other hand, the expression for the induced metric blows up when $\rho \to 0$. In this article, we will only use the rigged metric in our computations. 

We next introduce a notion of a connection on the space $H$, a \emph{rigged connection}, descended from the rigged structure. Recall that by definition, a rigged tensor field $T_a{}^b$ on $H$ is a tensor on $M$ such that $k^a T_a{}^b=0=T_a{}^b n_b$. We defined a rigged connection of a tensor field $T_a{}^b$ as a covariant derivative projected onto $TH$, 
\begin{align}
D_a T_b{}^c = \Pi_a{}^d \Pi_b{}^e (\nabla_d T_e{}^f)\Pi_f{}^c.
\end{align}
One first check that this connection is torsionless
\begin{equation}
\begin{aligned}
[D_a, D_b] F 
&= \Pi_a{}^c\Pi_b{}^d ( \nabla_c \Pi_d{}^e -  \nabla_d \Pi_c{}^e) \nabla_e F \\
&=-\Pi_a{}^c\Pi_b{}^d   (\nabla_c n_d - \nabla_d n_c) k[F] \\
&=0
\end{aligned}
\end{equation}
where we used in the last equality the fact that $n_a$ defines a foliation $\nabla_{[a}n_{b]}= a_{[a}n_{b]}$. It is also straightforward to check that the rigged connection preserves the rigged projector 
\begin{align}
D_a \Pi_b{}^c =   \Pi_a{}^d\Pi_b{}^e (\nabla_d \Pi_e{}^f) \Pi_f{}^c = -\Pi_a{}^d\Pi_b{}^e \nabla_d (n_e k^f) \Pi_f{}^c=0.
\end{align}
It does not, however, preserve the rigged metric and its conjugate. Instead, we can show that 
\begin{equation}
\begin{aligned}
D_a H^{bc} &=   \Pi_a{}^d \nabla_d (g^{ij} \Pi_i{}^e\Pi_j{}^f ) \Pi_e{}^b \Pi_f{}^c \\
&=   \Pi_a{}^d g^{ij}[\Pi_j{}^c (\nabla_d \Pi_i{}^e)  \Pi_e{}^b
+  \Pi_i{}^b ( \nabla_d \Pi_j{}^f )  \Pi_f{}^c ]\\
&=   -\Pi_a{}^d g^{ij}[ n_i  \Pi_j{}^c  (\nabla_d k^e)\Pi_e{}^b 
+  n_j \Pi_i{}^b ( \nabla_d k^f )  \Pi_f{}^c ]\\
&= - (K_a{}^b\ell^c  + K_a{}^c  \ell^b).
\end{aligned}
\end{equation}
where $K_a{}^b := \Pi_a{}^c(\nabla_c k^d)\Pi_d{}^b$ is the extrinsic curvature of the surface $H$ computed with the rigged metric. This tensor can be related to the rigged derivative of the tangent form $k_a$ as follows
\be \label{K}
K_a{}^c q_{ca} = (D_a +\omega_a)k_b,
\ee 
where $\omega_a:= \Pi_a{}^c (k^b\nabla_c n_b)$ is the rigged connection.

Given the rigged structure on the stretched horizon $H$ and a volume form $\bm{\epsilon}_M$ on the spacetime $M$ we can define the induced volume form on $H$ by the contraction, $\bm{\epsilon}_H := \iota_k \bm{\epsilon}_M$.
The conservation equation of this volume form involves the rigged connection as follows\footnote{This also means that $(D_a+ \omega_a) \bm{\epsilon}_H=0$.}
\be
\bm{\rd}(\iota_\xi \bm{\epsilon}_H) = [(D_a-\omega_a) \xi^a] \,\bm{\epsilon}_H. 
\ee
where $\xi$ is a vector tangent to $H$. 
Interestingly, this conservation equation can also be written in terms of the Carrollian structure as 
\be 
\bm{\rd}(\iota_\xi \bm{\epsilon}_H ) = 
\left[(\ell +\theta)[\tau] + (\sD_A +\varphi_A) X^A\right]\bm{\epsilon}_H, 
\ee
for a vector $\xi =\tau \ell + X^A e_A$.

\section{Conservation Laws on Stretched Horizons} \lb{sec:cons}

We are now at the stage where we can discuss Carrollian fluid energy-momentum tensor on the stretched horizon $H$ and derive its conservation laws. The plan is to outright define first the Carrollian fluid energy-momentum tensor and show how the Einstein equations imply conservation laws (or vice versa). The correspondence between fluid quantities and the extrinsic geometry of $H$, the so-called gravitational dictionary, will be discussed afterwards. 

Following the construction presented in \cite{Chandrasekaran:2021hxc}, the rigged energy-momentum tensor on the null boundary $N$ is related to the the null Weingarten tensor $\Pi_a{}^c \nabla_c \ell^d \Pi_d{}^b$. Since the vector $n^a$ goes to $\ell^a$ on $N$, it suggests that the fluid energy-momentum tensor on the timelike surface is defined as, 
\begin{align}
T_a{}^b = \sW_a{}^b -\sW \Pi_a{}^b, \lb{e-m-def}
\end{align} 
where the rigged Weingarten tensor (sometimes called the shape operator) on $H$ is defined to be\footnote{For the case $D_a \rho =0$ that we consider, the Weingarten tensor can be written simply as $\sW_a{}^b  = \Pi_a{}^c \nabla_c n^b$.}
\begin{align}
\sW_a{}^b := \Pi_a{}^c  (\nabla_c n^d) \Pi_d{}^b, \lb{wein}
\end{align}
and we denote its trace by $\sW = \sW_a{}^a$. Obviously, this rigged Weingarten tensor becomes the null Weingarten tensor \cite{Hopfmuller:2018fni, Chandrasekaran:2020wwn, Chandrasekaran:2021hxc} on the null boundary $N$. 
It captures essential elements of extrinsic geometry of the surface $H$ whose components have been established to serve as the conjugate momenta to the intrinsic geometry of the surface in the gravitational phase space (see \cite{Hopfmuller:2016scf,Hopfmuller:2018fni} for the case of null boundaries). In our construction, the intrinsic geometry of $H$ is encoded in the Carroll structure and, as we will explain later, the extrinsic geometry is the Carrollian fluid momenta. 
This energy-momentum tensor agrees on the null surface with the one defined in \cite{Chandrasekaran:2021hxc} on the null boundary except for the overall sign.  We will show next that the Einstein equations $G_{ab} =0$ and the condition $D_a \rho =0$, imply hydrodynamic conservation laws $D_b T_a{}^b =0$ .

\subsection{Conservation laws }

Our goal here is to show that conservation of energy-momentum tensor follows from the Einstein equations. In the following derivation, we will keep track of the tangential derivative of the norm of the normal form, $D_a \rho$, by allowing its value to be non-zero. We will show that the condition $D_a \rho =0$ is necessary to have a proper definition of energy-momentum tensor which obeys conservation laws outside the null boundary $N$, hence justifying our prior assumption. 

To start with, the covariant derivative of the vector $n$ decomposes as  
\begin{align}\label{Dn}
\nabla_a n^b  = \sW_a{}^b +(D_a \rho) k^b + n_a \nabla_k n^b, \qquad \text{and thus} \qquad \nabla_a n^a = \sW + k[\rho],
\end{align}
where we used that $n_a \nabla_b n^a = \frac{1}{2}\nabla_b (n_a n^a) = \nabla_b \rho$. The rigged covariant derivative of the rigged Weingarten tensor can then be written as
\begin{equation}
\begin{aligned}
D_b \sW_a{}^b &= \Pi_a{}^c  (\nabla_b\sW_c{}^d)\Pi_d{}^b  = \Pi_a{}^c \nabla_b \sW_c{}^b + \sW_a{}^c \nabla_k n_c. 
\end{aligned}
\end{equation}
We can then show that
\begin{equation}
\begin{aligned}
\Pi_a{}^c \nabla_b \nabla_c n^b &= \Pi_a{}^c \nabla_b (\sW_c{}^b + k^b D_c \rho + n_c \nabla_k n^b) \\
&= \Pi_a{}^c \nabla_b \sW_c{}^b + (D_a \rho) (\nabla_b k^b)  + \Pi_a{}^c \nabla_k (D_c \rho) + \Pi_a{}^c(\nabla_b n_c) (\nabla_k n^b) \\
&= D_b \sW_a{}^b + (D_a \rho) (\nabla_b k^b)  + \Pi_a{}^c \nabla_k (D_c \rho) + ( \Pi_a{}^c\nabla_b n_c - \sW_{ab}) \nabla_k n^b\\
&= D_b \sW_a{}^b + (D_a \rho) K_b{}^b  + \Pi_a{}^c \nabla_k (D_c \rho) - a_a k[\rho] , 
\end{aligned}
\end{equation}
where to arrive at the last equality, we defined $K_a{}^b:= \Pi_a{}^c \nabla_c k^b$ and we used the property that $\nabla_b k^b  = K_b{}^b - k^b\nabla_k n_b$,  and we also use that
\begin{equation}
\begin{aligned}
( \Pi_a{}^c\nabla_b n_c - \sW_{ab}) = \Pi_a{}^c ( \nabla_b n_c - \nabla_c n_d \Pi^d{}_b) &=
 \Pi_a{}^c ( \nabla_b n_c - \nabla_c n_d (\delta^d{}_b - n^d k_b)) \\
 &= \Pi_a{}^c ( a_b n_c-a_c n_b) + D_a \rho  k_b \\
 & =  -a_a n_b  + D_a \rho  k_b.
 \end{aligned}
 \end{equation}
Next, using the property that the Einstein tensor along the vector $n^a$ projected onto $H$ coincides with the Ricci tensor, $\Pi_a{}^c n^b G_{bc} = \Pi_a{}^c R_{nc}$, and invoking the definition of the Ricci tensor in term of the commutator, we derive
\begin{equation}
\begin{aligned}
\Pi_a{}^c G_{n c} = \Pi_a{}^c [\nabla_b, \nabla_c] n^b  &= \Pi_a{}^c \nabla_b \nabla_c n^b - D_a (\nabla_b n^b) \\
 & = D_b (\sW_a{}^b - \sW \Pi_a{}^b) + K_b{}^b D_a \rho - a_a k[\rho]+ \Pi_a{}^c [\nabla_k, D_c] \rho.  \\
\end{aligned}
\end{equation}
We then show that the last term can be manipulated as follows:
\begin{equation}
\begin{aligned}
\Pi_a{}^c [\nabla_k, D_c]  \rho&= 
 \Pi_a{}^c k^b (\nabla_b  \Pi_c{}^d)  \nabla _d \rho  -  \Pi_a{}^d  (\nabla _d k^b) \nabla_b \rho\\
&= -\Pi_a{}^c k^b (\nabla_b n_c{})  k [\rho] -  \Pi_a{}^d  n_b(\nabla _d k^b)  k[ \rho]  - \Pi_a{}^d  \nabla _d k^b D_b \rho\\
&=  \Pi_a{}^c  k^b \left( \nabla _c n_b    -  \nabla_b n_c{} \right) k[\rho]    - K_a{}^b D_b\rho \\
&=  a_a  k[\rho]  - K_a{}^b D_b\rho,
\end{aligned}
\end{equation}
where we used that $\nabla_{[a} n_{b]} = a_{[a} n_{b]}$ to arrive at the last equality. Finally putting everything together, the Einstein tensor can therefore be expressed as
\begin{align}
\Pi_a{}^c G_{n c} =  D_b \left( \sW_a{}^b  - \sW\Pi_a{}^b \right) -(K_a{}^b -K \Pi_a{}^b) D_b \rho. \label{conseq}
\end{align}
It is therefore clear that under the condition $D_a \rho =0$, the energy-momentum tensor \eqref{e-m-def} is conserved once imposing the Einstein equations $\Pi_a{}^c G_{n c} =0$,
\begin{equation}
\boxed{
\Pi_a{}^b G_{n b} =  D_b T_a{}^b =0.}  \lb{einstein-fluid}
\end{equation}
Remarks are in order here:
\begin{enumerate}[label = \roman*)]
\item To prove the conservation laws we have only used that the transverse vector $k$ is null. We didn't need to assume that $k$ is geodesic and affinely parameterized.

\item Conservation laws are automatically satisfied on the null boundary $N$ without posing an extra condition on $\rho$ as its value already vanishes on $N$. This again agrees with \cite{Chandrasekaran:2021hxc}.

\item We can check that the conservation equations \eqref{conseq} transform covariantly  under the rescaling symmetry $\delta_\epsilon (\ell, k)=(\epsilon \ell, -\epsilon k)$: This follows from the transformations of the  Weingarten and  extrinsic curvature 
\be 
\delta_\epsilon \sW_a{}^b = \epsilon \sW_a{}^b +\ell^b D_a\epsilon, \qquad 
\delta_\epsilon K_a{}^b = -\epsilon K_a{}^b, \qquad \delta_\epsilon\rho = 2\epsilon \rho.
\ee
And the use of the identity 
\be 
D_b[\ell^b D_a\epsilon- \ell[\epsilon]\Pi_a{}^b] = -(D_a\ell^b-D_c\ell^c \Pi_a{}^b)D_b\epsilon.
\ee

\item One can always reach the condition $D_a \rho =0$ by exploiting the fact that the rigging condition $n_a k^a =1$ only defines the normal form $\bm{n}$ and the transverse vector $k$ up to the rescaling $\bm{n} \to \Omega \bm{n}$, $\ell \to \Omega \ell $  and $k \to \Omega^{-1} k$ for a function $\Omega$ on $M$. We will comeback to this point again shortly.
\end{enumerate}

\subsection{Gravitational dictionary}

We have already defined the energy-momentum tensor of the stretched horizon $H$ and showed that it obeys conservation laws as desired. We now proceed to discuss the dictionary between gravitational degrees of freedom and Carrollian fluid quantities. First, as a tensor tangent to the stretched horizon $H$, the energy-momentum tensor decomposes in terms of the Carrollian fluid momenta \cite{Ciambelli:2018xat, Ciambelli:2018ojf, Freidel:2022bai} as
\begin{align}
T_a{}^b := \sW_a{}^b - \sW \Pi_a{}^b =  -  k_a \left( \sE \ell^b + \sJ^b \right)+ \pi_a \ell^b + \left( \sN_a{}^b + \sP q_a{}^b \right), \lb{e-m-gravity}
\end{align}
where its components are the fluid energy density $\sE$, the pressure $\sP$, the fluid momentum density $\pi_a$, the heat current $\sJ^a$, and the viscous stress tensor\footnote{$\sN_a{}^b$ can also be understood to be the finite distance analog of the news tensor.} $\sN_a{}^b= q_a{}^c(\nabla_c n^d)q_d{}^b$. The tensors $\pi_a, \sJ^a$ and $\sN_a{}^b$ are horizontal, meaning that we can express them as
\begin{align}
\pi_a = \pi_A e^A{}_a, \qquad \sJ^a = \sJ^A e_A{}^a, \qquad \text{and} \qquad \sN_a{}^b = \sN_A{}^B e^A{}_a e_B{}^b.
\end{align}
Let us also note that the viscous tensor is symmetric, $\sN_{AB} :=  q_{AB} \sN_B{}^C = \sN_{BA}$, and traceless, $\sN_A{}^A = 0$. It then follows from the definition of the energy-momentum tensor \eqref{e-m-gravity} that the Weingarten tensor \eqref{wein}, which is a tensor field on $H$, can be parameterized in terms of Carrollian fluid momenta as
\begin{align}
\sW_a{}^b = \sN_a{}^b + \frac{1}{2} \sE q_a{}^b + \pi_a \ell^b -  k_a \sJ^b - \left(\sP+\frac{1}{2}\sE \right) k_a \ell^b, \label{W0}
\end{align}
and the trace is $\sW = \frac12\sE - \sP$. 

We now spell out more precisely the expression of the horizon Carrollian fluid in terms of the gravitational extrinsic geometry of the stretched horizon $H$.
We find that since the vector $n^a$ is the linear combination of the tangential vector $\ell^a$ and the transverse vector $k^a$, the Weingarten tensor then decomposes as follows
\begin{align}
\sW_a{}^b =  D_a \ell^b + 2\rho K_a{}^b,
\end{align}
where we used that $\ell^a$ is tangent to $H$, so the first term is the rigged derivative of $\ell$ while the second term is proportional to 
 $K_a{}^b :=\Pi_a{}^c (\nabla_c k^d )\Pi_d{}^b$.
 In order to give the dictionary between the Carrollian fluid expressions and the gravitational entities we need to introduce the definition of the 
 extrinsic curvature tensors $\theta_{ab},\btheta_{ab}$, the Hájíček form $\pi_a$, the surface and vector accelerations $ (\kappa, A^a)$.
 These are defined below, as coefficient in the decomposition of $D_a\ell^b$ and $K_a{}^b$ and we find that 
\begin{align}
 D_a \ell^b & = \theta_a{}^b +  k_a A^b + \pi_a \ell^b  +\kappa k_a \ell^b \lb{D-l} \\
K_a{}^b &   = \btheta_a{}^b -k_a (\pi^b + \varphi^b).
\end{align}
Here all the vectors and tensors are tangential to the sphere distribution\footnote{ This means that $\ell^a \pi_a = k^a\pi_a=0$ and similarly for $(\theta_a{}^b, \btheta_a{}^b, A^a)$.}.
Note that the absence of the $\ell^b$ terms in $K_a{}^b$ is due to the fact that the vector $k$ is null. The surface acceleration and the momenta appears in the decomposition of the rigged connection\footnote{We can therefore express \eqref{D-l} similarly to \eqref{K} as 
\be 
(D_a -\omega_a)\ell^b = \theta_a{}^b +  k_a A^b.
\ee }(see sec \ref{sec:con})
\be 
\omega_a = \kappa k_a +\pi_a.
\ee 
The last term in the expression for $K_a{}^b$ simply follows from the evaluation
\be
\ell^a K_{ab}e_A{}^b= \iota_{e^A} D_\ell k = -\iota_{D_\ell e^A} k
=
-\iota_{[\ell, {e_A}]} k
- \iota_{D_{A}\ell} k =- \varphi_A -\pi_A.
\ee
In the next section we explore in more details the gravitational dictionary between Carrollian fluids and gravity.
\subsubsection{Viscous stress tensor and Energy density}
Let us first consider the spin-2 components of the rigged Weingarten tensor, which are the extrinsic curvature tensor, $q_a{}^c q_{bd} \sW_c{}^d = q_a{}^c q_b{}^d \nabla_c n_d$. Observe that this object is symmetric in its two indices which follows from the fact that the normal form $\bm{n}$ defines foliation, $\nabla_{[a} n_{b]} = a_{[a} n_{b]}$. Its trace corresponds to the Carrollian fluid energy density $\sE$, 
\begin{align}
\sE := q_a{}^b \nabla_b n^a  \qquad \text{or equivalently}, \qquad \sE := q^{AB} g(e_B, \nabla_{e_A} n), 
\end{align}
and the traceless part corresponds to the viscous stress tensor, $\sN_{ab} = \sN_{AB} e^A{}_a e^B{}_b$, of Carrollian fluids, 
\begin{align}
\sN_{ab} := q_{\langle a}{}^c q_{b \rangle}{}^d\nabla_c n_d, \qquad \text{or}, \qquad \sN_{AB} := g(e_B, \nabla_{e_A} n) - \frac{1}{2} q^{CD} g(e_D, \nabla_{e_C} n) q_{AB} . 
\end{align}

We can also define the expansion tensor\footnote{Note that  the tensor $D_a \ell^b$ does not truly describe the extrinsic geometry of the space $H$ as $\ell$ is tangent to $H$. Its values are completely determined by the intrinsic geometry, i.e. the Carrollian structure of the surface. } associated to the tangential vector $\ell$ to be $\theta_{ab} :=  q_a{}^c q_b{}^d \nabla_c \ell_d$. Components of this expansion tensor can be expressed in the horizontal basis as
\begin{align}
\theta_{AB} = g(e_B, \nabla_{e_A} \ell) = \frac{1}{2}\ell[q_{AB}] + \rho w_{AB}.
\end{align}
Interestingly, its anti-symmetric components are proportional to the Carrollian vorticity. The trace and the symmetric traceless components of tensor $\theta_{AB}$ are the expansion and the shear tensor associated with the tangential vector $\ell$, 
\begin{align}
\theta := q^{AB} \theta_{AB} = \ell[\ln \sqrt{q}], \qquad \text{and} \qquad \s_{AB} := \theta_{(AB)} - \frac{1}{2} \theta q_{AB}.  
\end{align}
In a similar manner, we define the extrinsic curvature tensor associated to the transverse direction $k$ as $\btheta_{ab} := q_a{}^c q_b{}^d \nabla_c k_d$, and its components can be expressed as 
\begin{align}
\btheta_{AB} = g(e_B, \nabla_{e_A} k) = \frac{1}{2}k[q_{AB}] - \frac{1}{2}w_{AB}. 
\end{align}
Observe that $\btheta_{AB}$ is not symmetric even on the null surface. Its trace and its symmetric traceless components are respectively the expansion and the shear associated to $k$ and they are given by
\begin{align}
\btheta := q^{AB} \btheta_{AB} = k[\ln \sqrt{q}], \qquad \text{and} \qquad \bar{\s}_{AB} := \btheta_{(AB)} - \frac{1}{2} \btheta q_{AB}.
\end{align}
Let us also note that the combination 
\begin{align}
g(e_B, \nabla_{e_A} n) = \theta_{AB} + 2\rho \btheta_{AB} = \frac{1}{2} n[q_{AB}]
\end{align}
is symmetric as we have already stated. The fluid energy density and the viscous stress tensor are given in terms of expansions and shear tensors by 
\begin{align}
\sE = \theta + 2\rho \btheta, \qquad \text{and} \qquad \sN_{AB} = \s_{AB} + 2\rho \bar{\s}_{AB}.  
\end{align}
It is important to appreciate that geometrically, the internal energy $\sE$ computes the expansion of the area element of the sphere $S$ along the vector $n$. On the null surface $N$, it therefore computes the expansion of the area element along null vector $\ell$, while the traceless part $\sN_{ab}$ corresponds to the shear tensor \cite{Hopfmuller:2016scf, Hopfmuller:2018fni,Chandrasekaran:2021hxc}. 
\subsubsection{Momentum density}
There are two spin-1 components of the energy-momentum tensor $T_a{}^b$. The first one corresponds to the Carrollian fluid momentum density, $\pi_a = \pi_A e^A{}_a$, which is defined as
\begin{align}
\pi_a := q_a{}^c k_b \nabla_c n^b, \qquad \text{or in the horizontal basis}, \qquad \pi_A := g(k, \nabla_{e_A} n). 
\end{align}
It then follows from the null rigged condition, $k^ak_a =0$, that $\pi_a = q_a{}^c k_b \nabla_c \ell^b$ is the Hájíček field computed with the basis vector $(\ell, k,e_A)$. The expression of the fluid momentum in terms of the Carrollian acceleration can be derived starting from the commutators (\ref{com-4}) as follows,
\begin{equation}
\begin{aligned}
e_A[\bar{\alpha}] = g(\ell, [k ,e_A]) &= g(\ell, \nabla_k e_A) - g(\ell, \nabla_{e_A} k)\\
& = g(k, \nabla_\ell e_A) + g(k ,\nabla_{e_A}\ell) \\
& = g(k, [\ell, e_A]) + 2 g(k ,\nabla_{e_A}\ell)\\
& = \varphi_A + 2\pi_A, \label{balpha}
\end{aligned}
\end{equation}
where to get from the first line to the second line, we repeatedly applied Leibniz rule and used that $g([k,\ell],e_A) =0$.  We therefore arrive at the expression for the fluid momentum in terms of the metric components
\begin{align}
\pi_A = \frac{1}{2} \left( e_A[\bar{\alpha}] - \varphi_A \right). 
\end{align}

\subsubsection{Carrollian heat current}
Another spin--1 quantity is the Carrollian heat current, $\sJ^a = \sJ^A e_A{}^a$, defined as
\begin{align}
\sJ^a := -q_b{}^a \nabla_\ell n^b, \qquad \text{or in the horizontal basis}, \qquad \sJ^A := - q^{AB} g(e_B, \nabla_\ell n)
\end{align}
This object is related to the tangential acceleration $A^a = q_b{}^a \nabla_\ell \ell^b$ of the vector $\ell$ and the Carrollian momentum density.
First we can evaluate the tangential acceleration as follows
\begin{align}\label{acc}
A_A = g(e_A, \nabla_\ell \ell ) = - g(\ell, [\ell, e_A]) - g(\ell, \nabla_{e_A} \ell) = (e_A+ 2 \varphi_A)[\rho].
\end{align} 
Observe that the acceleration vanishes on the null boundary $N$.
Then  one can check using \eqref{defnl} and repeatedly applying Leibniz rule and the commutators \eqref{com-4}, and the evaluation \eqref{balpha}, that
\begin{equation}
\lb{general-J}
\begin{aligned}
\sJ_A &= - g(e_A, \nabla_\ell \ell) - 2\rho g(e_A, \nabla_\ell k) \\
& = -A_A +2\rho g(e_A, [k,\ell]) +2\rho g(\ell, [k,e_A]) - 2\rho g(k, \nabla_{e_A} \ell) \\
& = -A_A +2\rho \left( e_A[\bar{\alpha}] -\pi_A \right)\\
& = (-e_A +2\pi_A)[\rho].
\end{aligned}
\end{equation}
This Carrollian current also vanishes on the null boundary $N$. 

For the choice of null vector that keeps $\rho$ constant on the stretched horizon $H$, we simply have that 
\begin{align}
\sJ_A = 2\rho \pi_A, \qquad \text{and}\qquad A_A = 2\rho \varphi_A. \lb{J&A}
\end{align}

\subsubsection{Surface gravity and Pressure}
The last spin-0 component of the energy-momentum tensor is the fluid pressure $\sP$ defined as the combination 
\begin{align}
\sP = -\left[\kappa + \frac{1}{2} (\theta + 2\rho \btheta)\right].
\end{align}
$\sP$ is  the generalization of what is called the gravitational pressure in \cite{Hopfmuller:2018fni} defined for the case of null boundary. The surface gravity $\kappa$ is defined as\footnote{ We have that $\kappa  = g(k,\nabla_\ell n)$.} 
\begin{align}
\kappa = k_a \nabla_\ell \ell^a = g(k,\nabla_\ell \ell).
\end{align}
It measures the vertical acceleration of the vector $\ell$. Its value is non-zero even on the null boundary $N$. Let us also comment that we write the directional derivative of the Carrollian vector field $\ell$ along itself as 
\begin{align}
\nabla_\ell \ell = \kappa \ell + A^A e_A - (\ell-2\kappa)[\rho] k \Neq \kappa \ell,
\end{align}
Recalling that $A^A \Neq 0$, this means $\nabla_\ell \ell = \kappa \ell$ which clearly dictates that on the null boundary $N$, the Carrollian vector $\ell$ generates non-affine null geodesics, and the in-affinity is measured by the surface gravity $\kappa$. We can show that the surface gravity is given by
\begin{align}
\kappa = g(k,\nabla_\ell \ell) = - g(\ell, [\ell, k]) - g(\ell, \nabla_k \ell)  = \ell[\bar{\alpha}] + (k+2\bar{\kappa})[\rho].  \lb{kappa}
\end{align}
Let us additionally note that the inaffinity $\bar{\kappa}$ of the null geodesics generated by the rigging vector $k$ can be computed directly from the commutator $[k,\ell]$ provided in \eqref{com-4} and it is given in coordinates by 
\begin{align}
\bar{\kappa} = k[\alpha].
\end{align}

Let us summarize below the dictionary between Carrollian fluid quantities and the gravitational entities given by the components of the Weingarten tensors: In the frame where $D_a\rho=0$, we have that 
\begin{tcolorbox}[colback=white]
\vspace{-15pt}
\begin{subequations}
\lb{dictionary}
\begin{flalign}
&\text{\bf Energy density:} &\mathscr{E} & = \theta + 2\rho \btheta && \lb{energy}  \\
&\text{\bf Pressure:} &\mathscr{P} & = - (\ell[\bar{\alpha}] + (k+2\bar{\kappa})[\rho]) -\frac12(\theta + 2\rho \btheta)  \lb{pressure}\\
&\text{\bf Momentum density:} &\pi_A & = \frac{1}{2} \left( e_A[\bar{\alpha}] - \varphi_A \right), \lb{momentum} \\ 
&\text{\bf Carrollian heat current:} &\mathscr{J}^A & =  2\rho \pi_A,  \lb{current}\\
&\text{\bf Viscous stress tensor:} &\sN_{AB} & =   \s_{AB} + 2 \rho \bar{\s}_{AB}. \lb{stress}
\end{flalign}
\end{subequations}
\end{tcolorbox}
Note also that the Weingarten tensor can be written in a compact manner in terms of the gravitational data as
\be 
\sW_a{}^b  = (\theta_a{}^b +2 \rho \btheta_a{}^b) + \pi_a \ell^b + 2\rho k_a \pi^b  +\kappa k_a \ell^b .
\ee

Lastly and for completeness, let us provide the form of the covariant derivative of the normal vector $n = \ell + 2\rho k$ along $k$. This expression which enters the development of the normal derivative \eqref{Dn},  becomes handy in further computations,\footnote{ We  use that \be
g(e_A, \nabla_k n)= g(e_A, \nabla_k \ell)= -  g( \nabla_k e_A,  \ell)= -g(\nabla_\ell {e_A}, k )=
-g([\ell, {e_A}], k )-g(\nabla_{e_A}\ell , k)=-(\varphi_A +\pi_A).
\ee
}
\begin{align}
\nabla_k n^b &=  k[\rho]k^b - (\pi^b + \varphi^b) - \bar\kappa \ell^b.
\end{align}

\subsection{Rigged derivative summary}
It is now a good place for us to summarize our finding and write the expansion of the rigged derivative in terms of tangential entities. We have found that the rigged structure defines on the stretched horizon $H$ a rigged connection $D_a$ (which can be equivalently called a Carrollian connection) and a volume form $\bm{\epsilon}_H$. 
The compatibility of this rigged derivative and the volume form gives $(D_a+\omega_a)\bm{\epsilon}_H=0$, where we recall that $\omega_a =\kappa k_a + \pi_a$. We also have 
\begin{align}
(D_a-\omega_a) \ell^b & = \theta_a{}^b +   k_a A^b ,\\
(D_a +\omega_a) k_b &= \btheta_{ab}  - k_a (\pi_b + \varphi_b).
\end{align}
An important remark is that when the rigged connection preserve the vertical direction, $(D_a - \omega_a)\ell^b =0$, which means both the expansion $\theta_{ab}$ and the acceleration $A^a$ have to vanish, it defines a Carroll G-structure (or a strong Carroll structure) \cite{Bekaert:2015xua,Figueroa-OFarrill:2019sex,Figueroa-OFarrill:2020gpr, Herfray:2021qmp, Ashtekar:2021wld,Baiguera:2022lsw, Hansen:2021fxi}. 
The derivative of the tangential projector is expressed simply in terms of these tensors as 
\be 
D_a q_c{}^b   = - [(D_a+\omega_a) k_c]\ell^b - k_c [(D_a-\omega_a) \ell^b].
\ee 
We can also evaluate the derivative of the frame and its inverse as 
\begin{align}
D_a e_A{}^b
&= \two\Gamma^b_{aA}  - [\btheta_{aA} -k_a (\pi_A + \varphi_A) ] \ell^b + k_a \theta_A{}^b,\\
D_a e^A{}_b & =- \two\Gamma^A_{ab}  - (\theta_a{}^A + k_a A^A)k_b - k_a \theta_b{}^A, 
\end{align}
where we use the obvious notation 
$\theta_a{}^B= \theta_a{}^b e^B{}_b$ and $\two\Gamma^b_{aA} = e^A{}_a \two\Gamma^C_{BA} e_C{}^b$ where $\sD_{e_A}e_B=\two\Gamma_{AB}^C e_C$ are the component of the horizontal connection.
This shows that the rigged derivative depends on the components of the rigged connection $(\kappa, \pi_a)$ and on the kinematical Carrollian elements such as the Carrollian acceleration and vorticity $(\varphi_a,w_{ab})$. It also contains elements which are intrinsic such as the expansion tensor $\theta_{(ab)}=\frac12 \cL_{\ell} q_{ab}$. Finally, it contains also extrinsic elements such as the extrinsic curvature $\btheta_{(ab)}$ that we refer to as the \emph{shear}\footnote{ As we have seen the anti-symmetric components of the extrinsic tensor is given by the Carrollian vorticity $\btheta_{[ab]}=\frac12 w_{[ab]}$.}, the acceleration $A^a$ and the anti-symmetric components of the expansion tensor. 
When the rigged connection is derived from an embedding, we have that the acceleration can be expressed as $A_a = \sD_a\rho +2\rho\varphi_a$, where $2\rho$ is the norm of the normal vector. It also means that the anti-symmetric components of the expansion tensor $\theta_{[ab]}=\frac{\rho}2 w_{ab}$ is proportional to the Carrollian vorticity. In other words, the rigged connection derived from a rigged structure depends on the metric $q$ but also on $(\rho,\omega_a)$ and the shear $\btheta_{ab}$. The shear tensor can be understood as encoding the gravitational radiation of the stretched horizon $H$.


\subsection{Comment on the energy-momentum tensor}
As we have explained, the condition $D_a \rho=0$ is necessary to have conservation of the energy-momentum tensor \eqref{e-m-gravity} and that this condition can always be chosen by properly rescaling the normal form $\bm{n}$. Let us now demonstrate how this is done. Suppose that we start from a normal $\wh{\bm{n}}$ with norm $2\hat{\rho}$ that is not constant on the surface, $D_a \hat{\rho} \neq 0$, and consequently the energy-momentum tensor 
$\wh{T}_a{}^b$ naively defined as in \eqref{e-m-gravity}, with $\wh{\bm{n}}$ replacing $\bm{n}$, is no longer conserved. 
\begin{align}
\wh{T}_a{}^b := \wh{\sW}_a{}^b - \wh{\sW} \Pi_a{}^b =  - \left( \wh{\sE} \ell^b + \wh{\sJ}^b \right) k_a + \wh{\pi}_a \ell^b + \left( \sN_a{}^b + \wh{\sP} q_a{}^b \right), \lb{e-m-gravity-2}
\end{align}
where $\wh{\sW}_a{}^b$ is the Weingarten tensor now defined with the rescaled vector $\wh{n}^a$.

In close vicinity of the null boundary $N$, we can always express the norm as $\hat{\rho} = r \eta$, where $\eta$ is a strictly positive function on $M$. We can now define the new normal form as
\begin{align}
  \bm{n} := \frac{1}{\sqrt{\eta}} \wh{\bm{n}}, \qquad \text{with its norm being} \qquad n_a n^a = 2r,
\end{align}
which is now constant on the surface $H$. Notice that this corresponds to the change in the scale factor $\hat{\bar{\alpha}}\to \bar{\alpha}= \hat{\bar{\alpha}} - \frac12 \ln \eta$. 
 The conserved  energy-momentum tensor $D_b T_a{}^b =0$ is the one defined in terms of $n$. One can check that this new conserved tensor is related to the naive, non-conserved, one by
\begin{align}
T_a{}^b= \frac{1}{\sqrt{\eta}} \left( \wh{T}_a{}^b - q_a{}^c \pa_c (\ln\sqrt{\eta}) \ell^b + \ell[\ln \sqrt{\eta}] q_a{}^b \right).  
\end{align}
Note that when working with the closed normal form $\wh{n}=\rd r$, such that  $\hat{\bar{\alpha}} =0$, the function $\eta$ coincides, on the null boundary, with the surface gravity $\hat{\kappa}$ of $\hat{\ell}$. In such case, this particular form of the conserved energy-momentum $\wh{T}_a{}^b$, with the presence of the derivatives $D_a \ln \sqrt{\kappa}$ terms, has been proposed in \cite{Donnay:2019jiz}. 
In our previous construction, we have already bypassed this construction by assuming a priori the condition $D_a\rho = 0$.

\subsection{Einstein equations on the stretched horizons}

We have already proved the the Einstein equations corresponds to the conservation laws of energy-momentum tensor \eqref{e-m-gravity}. With the extrinsic geometry of the stretched horizon $H$ defined, we now finally explicitly write the Einstein equations on $H$ in terms of the Carrollian fluid momenta. 

Following from the conservation equation \eqref{einstein-fluid}, the component $G_{n \ell}$ of the Einstein tensor can be written, by recalling the definition of the energy-momentum tensor \eqref{e-m-gravity} and the rigged covariant derivative \eqref{D-l}, as   
\begin{equation}
\begin{aligned}
G_{n \ell} & = \ell^a D_b T_a{}^b   \\
&= D_a (\ell^b T_b{}^a) - T_a{}^b D_b \ell^a\\
& = -D_a\left( \sE \ell^a +\sJ^a \right) - T_b{}^a \left(  \theta_a{}^b + \pi_a \ell^b + A^bk_a + \kappa k_a \ell^b \right)  \\
& = -(\ell + \theta)[\sE] - \sP \theta -  (\sD_A + \varphi_A) \sJ^A -A_A\pi^A  - \sN_A{}^B \theta_B{}^A  \\
&= -(\ell + \theta)[\sE] - \sP \theta -(\sD_A + 2\varphi_A)\sJ^A - \sN_A{}^B \sigma_B{}^A,\label{Econs}
\end{aligned}
\end{equation}
where we used that $D_a \sJ^a  = \sD_A \sJ^A + (\pi_A + \varphi_A) \sJ^A$ and $D_a \ell^a = \theta + \kappa$ (derivations are given in Appendix \ref{app:derivative}), and to obtain the last equality we also used that $A_A \pi^A = \varphi_A \sJ^A$ that follows from \eqref{J&A}. The remaining components of the Einstein tensor are $G_{n A}$, which in a similar manner, we can use the energy-momentum tensor \eqref{e-m-gravity} and the rigged derivative of the horizon basis, $D_b e_A{}^a$, provided in \eqref{D-e_A} to show that
\begin{equation}
\begin{aligned}
G_{n A} & = e_A{}^a D_b T_a{}^b   \\
&= D_a (e_A{}^b T_b{}^a) - T_a{}^b D_b e_A{}^a\\
& = D_a\left( \sN_A{}^a + \sP e_A{}^a+ \pi_A \ell^a \right) -\two\Gamma^C_{BA} \sN_C{}^B - \pi^B \theta_{AB} - \sJ^B\btheta_{BA}  + \sE(\pi_A + \varphi_A)\\
&= (\ell +\theta + \kappa )[\pi_A]+(D_B + \pi_B + \varphi_B) (\sN_A{}^B + \sP \delta_A^B)- \pi^B \theta_{AB} - \sJ^B\btheta_{BA}  + \sE(\pi_A + \varphi_A)\\
&= (\ell + \theta)[\pi_A] + \sE \varphi_A- w_{AB}\sJ^B + (\sD_B + \varphi_B) (\sN_A{}^B +\sP \delta_A^B), \label{Mcons}
\end{aligned}
\end{equation}
where we used again that $D_a e_A^a = \two\Gamma^B_{BA} + (\pi_A +\varphi_A)$ and $D_a \ell^a = \theta + \kappa$ (see Appendix \ref{app:derivative} for explanations), and to obtain the last equality, we utilized the gravitational dictionary \eqref{dictionary}, more specifically the following relations: $\theta_{AB}+ 2\rho \btheta_{AB} = \sN_{AB}  +\frac{1}{2}\sE q_{AB}$, $w_{AB} = \btheta_{BA} - \btheta_{AB}$, $2\rho\pi^A = \sJ^A$, and $\sP = - \kappa - \frac{1}{2}\sE$.  
This shows that the vaccuum Einstein's equation projected on stretched horizons
 are Carrollian fluid conservation equation \cite{Ciambelli:2018xat, Donnay:2019jiz, Freidel:2022bai}. 
The conservation equations are \eqref{Econs} and \eqref{Mcons} are Carrollian fluid conservation equations. These can be conveniently written as 
\bea
\ell[\sE] + (\sP+\sE) \theta &=& -(\sD_A + 2\varphi_A)\sJ^A - \sN_A{}^B \sigma_B{}^A, \\
(\ell + \theta)[\pi_A] + (\sE+\sP) \varphi_A +\sD_A \sP &=&  w_{AB}\sJ^B - (\sD_B + \varphi_B) \sN_A{}^B.
\eea 
where the RHS represents fluid dissipation effects. The null Carrollian fluid equations are recovered when $\sJ^A=0$.

\subsection{Einstein Equations on the null boundary}
In the previous section we have shown that  the Einstein equation $G_{na}$ projected on the stretched horizon $H$ are equivalent to conservation equations. These equations are independent of the shear of the Carrollian connection. Ultimately it is essential to look at the rest of the Einstein's equation projected on $H$. Here, we achieve this but only for the Einsteins equations projected onto the null surface $N$. We denote with a ring the projected tensors: $\mr{q}_{AB}, \mr{\theta}_{AB},\cdots$ denote the evaluation of $q_{AB}, \theta_{AB}, \cdots $ onto $N$.

We find that the  components of the Einstein tensor on the null boundary are 
\begin{tcolorbox}[colback=white]
\vspace{-15pt}
\begin{subequations}
\begin{align}
-\mr{G}_{\ell \ell} &= (\ell + \mr{\theta})[ \mr{\sE}] + \mr{\sP} \mr{\theta} + \mr{\s}_A{}^{B} \mr{\s}_B{}^{A} \\
\mr{G}_{\ell A} & = (\ell + \mr{\theta} )[\pi_A]+ (\mr{\sE}+\mr{\sP}) \varphi_A+(\mrD_B+\varphi_B) \mr{\s}_A{}^B  \\
\mr{G}_{\ell k} 
&=(\ell + \tfrac12\mr{\theta} -\sP )[ \btheta]  + (\mrD_A + \pi_A + \varphi_A) (\pi^A+\varphi^A)   - \frac{1}{2}\two R \label{Glk}\\
- \mr{G}_{\langle A B \rangle} &= \left[ 2(\ell-\mr{\theta} - \mr{\sP}) [\bar{\s}_{AB}]  +\btheta\mr{\s}_{AB} + 2( \mrD_A + \pi_A +\varphi_A) (\pi_B+\varphi_B) \right]_{\mathrm{STF}}
\end{align}
\end{subequations}
\end{tcolorbox}
The subscript STF means that we take the symmetric trace free components\footnote{In particular we have that 
\be
\left[\ell[ \bar\sigma_{AB}]\right]^{\mathrm{STF}} = 
\ell[ \bar\sigma_{AB}] - 2  \bar{\s}_{C( A} \mr{\s}_{B)}{}^C =  \ell[ \bar\sigma_{AB}] - q_{AB} (\bar{\s}_{C}{}^D \mr{\s}_{D}{}^C)
\ee.}.
The first two equations are simply the null Carrollian conservation equation we have just derived and they are known as the null Raychaudhuri equation and the Damour equations, respectively. 
The next two equations determine the evolution of the shear $\btheta_{AB}$ in terms of $\sP$ and the intrinsic geometry data $(q_{AB},\varphi_A,\pi_A,\theta_{AB})$.
It is important to check that these equations are invariant under the rescaling symmetry
\bea
(\ell,
\mr{\theta}, \mr{\sE},\sN_A{}^B) &\to& (e^\epsilon \ell, e^\epsilon\mr{\theta}, e^\epsilon \mr{\sE},e^{\epsilon}\sN_A{}^B) \\
\mr\sP \to e^\epsilon (\mr \sP -\ell[\epsilon]),
\quad
\varphi_A &\to& \varphi_A -e_A[\epsilon], 
\quad 
\pi_A \to \pi_A + e_A[\epsilon].
\eea
 In addition, the other Einstein equations involve the trace part of the components $\mr{G}_{A B}$. In the gauge where $\bar\kappa=0$, i.e., where $k$ is affinely parameterized, it is given by,\footnote{ Equating \eqref{Glk} with \eqref{trG} means that 
\be 
k[\kappa] + (\pi+\varphi)\!\cdot\!(\pi+\varphi) = \frac12 \theta \btheta - \sigma:\bar\sigma -\tfrac12 \two R
\ee} 
\begin{align}\label{trG}
\frac{1}{2}\mrq^{AB} \mr{G}_{AB} = - R_{\ell k} = (\ell - \mr{\sP}) [\btheta]  + k[\kappa] +(\mrD_A + 2(\pi_A +\varphi_A) ) (\pi^A+\varphi^A) + \mr{\s}:\bar{\s}. 
\end{align}
A more detailed study of these equations and their interpretation in terms of symmetries will be performed in \cite{forthcoming}.

\section{Symmetries and Einstein Equations} \lb{sec:symplectic}

The last part of this work aims at exploring the gravitational phase space, symmetries, and the associated Noether charges of the stretched horizon. We would like to demonstrate the following points:

\begin{enumerate}[label = \roman*)]
\item The pre-symplectic potential of the gravitational phase space of the stretched horizon $H$ is essentially expressed in terms of the Carrollian conjugate pairs as in \cite{Freidel:2022bai}. 

\item The tangential components of the Einstein equations, namely $\Pi_a{}^b G_{n b} =0$, which are interpreted Carrollian hydrodynamics, are conservation equations derived from  the diffeomorphism symmetries of the stretched horizon. We will also compute the Noether charges associated with these symmetries. 
\end{enumerate}



\subsection{Pre-Symplectic Potential of stretch horizons}

Gravitational phase space of the stretched horizon $H$ can be constructed using the covariant phase space formalism. Following the covariant phase space formalism, we look at the \emph{pre-symplectic potential} that encodes the phase space information of the theory. In this study, we consider the 4-dimensional Einstein-Hilbert Lagrangian without the cosmological constant term and without matter degrees of freedom, meaning that $\bm{L} = \frac{1}{2} R \bm{\epsilon}_M$ where $R$ is the spacetime Ricci scalar and $\bm{\epsilon}_M$ denotes the spacetime volume form. The standard pre-symplectic potential of the Einstein-Hilbert gravity pulling back to the stretched horizon $H$ is given by
\begin{equation}
\begin{aligned}
\bm{\Theta}_H = -\Theta^a n_a \bm{\epsilon}_H, \qquad \text{where} \qquad  \Theta^a = \frac{1}{2} \left( g^{ac} \nabla^b \delta g_{bc} - \nabla^a \delta g \right),
\end{aligned}
\end{equation}
where we recalled the volume form on the surface $\bm{\epsilon}_H := -\iota_k \bm{\epsilon}_M$ and we also denoted the trace of the metric variation with $\delta g := g^{ab} \delta g_{ab}$. 

To evaluate the pre-symplectic potential $\bm{\Theta}_H$, one starts with the variation of the spacetime metric, whose components can be expressed in terms of the co-frame fields as,
\begin{align}
\delta g_{ab} = \delta q_{ab}   + 2k_{(a} \delta n_{b)} + 2\ell_{(a} \delta k_{b)} - 2(\delta \rho) k_a k_b.
\end{align}
Computations of the variation $\delta g_{ab}$ thus boils down to the computation of variations of the co-frame $\bm{n}$ and $\bm{k}$ and the null metric $q_{ab}$. These variations are given by\footnote{We also have the field variation of the Carrollian vector
\be
\delta \ell = -\bbdelta \alpha \ell +\e^{-\alpha} \bbdelta V^A e_A.\label{varell}
\ee}
\begin{align}
\delta \bm{n} = \delta \bar{\alpha} \bm{n}, \ \ \ \delta \bm{k} = \bbdelta \alpha \bm{k} - \e^\a \bbdelta \beta_A \bm{e}^A, \ \ \ \delta q = - 2 \e^\a q_{AB}\bbdelta V^B \bm{k} \odot \bm{e}^A + \bbdelta q_{AB} \bm{e}^A \odot \bm{e}^B, \lb{var-form}
\end{align}
where we define the variation $\bbdelta$ as follows
\begin{align}
\bbdelta\a &:= \delta \a + \beta_A \bbdelta V^A , \\
\bbdelta\beta_A & : =  (J^{-1})_A{}^C\delta \left( J_C{}^B\beta_B \right) - (\beta \cdot \bbdelta V)\beta_A, \\
\bbdelta q_{AB} &:= (J^{-1})_A{}^C (J^{-1})_B{}^D\delta \left(J_C{}^E J_D{}^F q_{EF}\right) - 2q_{C(A} \beta_{B)} \bbdelta V^C, \\
\bbdelta V^A &:= \left(\delta V^B\right)J_B{}^A. 
\end{align}
These field variations can also be written in terms of the variations of the fundamental forms and vectors as
\be 
\delta\bar\a = k^a \delta n_a, \qquad \bbdelta\a = \ell^a \delta k_a,
\qquad  \e^{\a} \bbdelta\beta_A =-  e_A{}^a \delta k_a, \qquad \e^{-\alpha} \bbdelta V^A= e^A{}_a \delta \ell^a
\ee 
One can then compute the trace of the metric variations and it is given by
\begin{align}
\delta g = 2(\delta \bar{\alpha} +\delta \alpha + \delta \ln \sqrt{q})  = 2(\delta \bar{\alpha} +\bbdelta \alpha + \bbdelta \ln \sqrt{q}).
\end{align}

After tedious but straightforward computations (see the derivation in section \ref{potential-derive}), we finally obtain the expression for the pre-symplectic potential on the stretched horizon. This potential is the sum of three terms: a bulk canonical term, a total variation term, and a boundary term as $\Theta_H= \Theta_H^{\mathrm{can}} +\delta L_H + \Theta_S$, where each term is

\begin{tcolorbox}[colback=white, coltitle=white, left=5pt, right=2 pt]
\vspace{-15pt}
\begin{subequations}
\label{potential}
\begin{flalign}
\Theta_H^{\mathrm{can}} & = \int_H \left( -\sE \bbdelta \alpha +   \e^\a \sJ^A \bbdelta \beta_A - \pi_A \e^{-\a} \bbdelta V^A  + \frac{1}{2}\left( \sN^{AB} +\sP q^{AB} \right)\bbdelta q_{AB} -\btheta \delta \rho \right) \bm{\epsilon}_H && \lb{potential-can} \\
L_H & = \int_H \left( \kappa + \sE \right) \bm{\epsilon}_H && \\
\Theta_S & = \frac{1}{2}\int_S  \delta\left(\alpha -  \bar{\alpha} \right)\bm{\epsilon}_S &&
\end{flalign}
\end{subequations}
\end{tcolorbox}
\noindent Note that we can use the identity $\int_H \theta \bm{\epsilon}_H= \int_S \bm{\epsilon}_S$ to rewrite part of the second term as a corner term. We first observe that the bulk canonical piece of the pre-symplectic potential contains the same conjugate pairs as in the action for Carrollian hydrodynamics \cite{Freidel:2022bai} with the addition of the term $\btheta \delta \rho$ that vanishes on the null boundary $N$. We also notice that the scale $\bar{\alpha}$ of the normal form only appear in the corner term, in agreement with the one presented in \cite{Hopfmuller:2016scf, Hopfmuller:2018fni} for the case of null boundary. This suggests that we can safely set $\bar{\alpha} =0$ without losing any phase space data. Let us mention \cite{Parattu:2016trq} for an earlier unified description of null and timelike pre-symplectic structure.

An important check that this expression \eqref{potential} is the right one comes from checking the fact that it is invariant under the rescaling transformation \eqref{rescaling}. The infinitesimal rescaling $\delta_\epsilon \ell = -\epsilon \ell$, $\delta_\epsilon k = \epsilon k$ implies the following transformations
\bea
\delta_\epsilon(\sE \bm{\epsilon}_H) =0, \quad \delta_\epsilon(\kappa\bm{\epsilon}_H)=  -\ell[\epsilon]\bm{\epsilon}_H, \quad \delta_\epsilon \rho = -2 \epsilon \rho, \quad 
\delta_\epsilon \alpha=-\delta_\epsilon\bar\alpha  =\epsilon, \quad \delta_\epsilon \pi_A = - e_A[\epsilon].
\eea
We can then check that 
\bea
\delta_\epsilon \Theta^{\mathrm{can}}_H &=&
\int_H 
\left(- \sE \delta \epsilon + \e^{-\alpha}\bbdelta V^A e_A[\epsilon] +  \frac{1}{2}\ell[\epsilon] q^{AB}\delta q_{AB} +2 \rho \btheta \delta \epsilon \right) \bm{\epsilon}_H \\
&=& \int_H \bigg(
\left(
(- \sE + 2\rho \bar\theta) \delta \epsilon 
+ (\delta \ell)[\epsilon]\right) 
\bm{\epsilon}_H + 
\ell[\epsilon] (\delta \bm{\epsilon}_H) \bigg)
  \\
  &=& -\int_H (\ell+\theta)[ \delta \epsilon]  \bm{\epsilon}_H + \delta\left( \int_H  \ell[\epsilon]  \bm{\epsilon}_H \right)
\\
&=& -\int_S  (\delta \epsilon)   \bm{\epsilon}_S + \delta\left( \int_H  \ell[\epsilon]\bm{\epsilon}_H \right),
\eea
where in the second equality we used \eqref{varell} and the variation 
$\delta \bm\epsilon_H =(\delta \alpha + \tfrac12 q^{AB} \delta q_{AB}) \bm\epsilon_H $.
From this we see that 
$
\delta_\epsilon \Theta_H =0,
$
inferring the invariance of the pre-symplectic potential under the rescaling. This implies that 
\be 
I_\epsilon \Omega_H = \delta_\epsilon \Theta_H -\delta (I_\epsilon \Theta_H) =-\delta (I_\epsilon \Theta_H).
\ee The corresponding canonical charge is therefore
\be 
I_\xi \Theta_H = \int_H (-\sE \epsilon + 2 \rho \bar\theta \epsilon)\bm{\epsilon}_H - \int_H \ell[\epsilon]\bm\epsilon_H +\int_S \epsilon \bm{\epsilon}_S =0.
\ee Since it vanishes, this means that the rescaling is indeed a gauge symmetry.

Using the same strategy, we can prove that the boost symmetry \eqref{boost} is a gauge symmetry provided we impose the condition 
\be 
\sJ_A= 2\rho\pi_A.
\ee

\subsection{Noether Charges for tangential symmetries} 
We now show that the pre-symplectic potential we have just described is symmetric under diffeomorphism tangent to the stretch horizon $H$,
\be
\xi= \tau \ell + X, \qquad \text{where} \qquad X:= X^A e_A.
\ee
The transformation rules for the metric coefficients are easily determined by demanding that the transformation rules of the fundamental forms and vectors $(\ell, \bm{k}, \bm{n}, g)$ are non-anomalous\footnote{This means that $\Delta_\xi (\ell, \bm{k}, \bm{n}, g)=0$ with the (field independent) anomaly operator defined as the difference between the field variation and the Lie derivative, $\Delta_\xi =\delta_\xi -\cL_\xi$. More details and applications related to this technology can be found in \cite{Hopfmuller:2018fni,Chandrasekaran:2020wwn,Chandrasekaran:2021hxc,Freidel:2021fxf,Freidel:2021cjp,Freidel:2021dxw}.},
This means in particular that  
one first has to write down the transformation rules for relevant basis vectors and 1-forms.
These are given by 
\bea
\cL_\xi \bm{k} &=& (k+\bar\kappa)[\tau] \bm{n} + \left(\ell[\tau] +X^A\varphi_A \right) \bm{k} 
+ \left((e_A-\varphi_A)[\tau] +w_{AB} X^B  \right) e^A,\\
\cL_\xi \bm{n} &=& \xi[\bar\alpha] \bm{n}, \\
\cL_\xi \ell &=& -(\ell[\tau] + X^A \varphi_A) \ell - \ell[X^A] e_A.
\eea  
One remark is that, demanding that the diffeomorphism $\xi$ preserves the condition that the Ehresmann connection $\bm{k}$ is tangent to the horizon $H$ requires that  $(k+\bar\kappa)[\tau]=0$.
We assume that this condition is satisfied.
Following from \eqref{var-form} and \eqref{varell} the transformation rules 
\bea
\delta_\xi\bar\alpha &=& \xi[\bar\alpha],\cr
\bbdelta_\xi \alpha &=& \ell[\tau] +X^A\varphi_A,  \cr 
-\e^\alpha \bbdelta_\xi\beta_A &=&  (e_A-\varphi_A)[\tau] +w_{AB} X^B,\cr 
-\e^{-\alpha} \bbdelta_\xi V^A &=& \ell[X^A],\cr 
\bbdelta_\xi q_{AB} &=& 2\left(\tau \theta_{AB} + \sD_{(A} X_{B)} \right),\cr 
\delta_\xi \rho &=& \xi[\rho].
\eea 
We then evaluate that 
\begin{align}\label{sympcan}
I_{\xi} \Theta^{\text{can}}_H = -\int_H \left(\tau {G}_{ n \ell} +Y^A {G}_{n A}\right) \bm{\epsilon}_H + {Q}_{(\tau,Y)}. 
\end{align}
We now see that the stretched horizon $H$ Raychaudhuri equation ${G}_{n \ell} =0$ and the Damour equations $G_{n A}$ are associated with the tangential diffeomorphism on $H$. 
This extends to the stretched horizon what has already been established in the literature for null surfaces (see \cite{Hopfmuller:2018fni}). The Noether charges are given (for non-zero $\beta_A$) by
\begin{align}
{Q}_{(\tau,Y)} = \int_S \left(  -\tau {\sE} + Y^A \left(\pi_A + ({\sN}_A{}^B + {\sP} \delta_A^B) \e^\a \beta_B\right)\right) \bm{\epsilon}_S.
\end{align}
They are precisely the charges for Carrollian hydrodynamics \cite{Ciambelli:2018ojf,Freidel:2022bai}.

\subsection{Covariant derivation of the Einstein equations}
For completeness, we provide here a detailed derivation of \eqref{sympcan} using the covariant form of the pre-symplectic potential.
First we use that we can write the bulk canonical pre-symplectic potential $\Theta^{\text{can}}_H$ \eqref{potential-can} in a covariant manner as\footnote{
Note that the variational term contracting $T_a{}^b$ can be written 
\bea
  \Pi_b{}^c \ell^a \delta k_c 
+ q^{ac}\left(-  k_b \delta \ell^d q_{cd}
+\frac{1}{2} q_b{}^d \delta q_{cd} \right)
= \Pi_b{}^c \ell^a \delta k_c 
+ q^{ac}\left(  k_b  \ell^d  \delta q_{cd}
+\frac{1}{2} q_b{}^d \delta q_{cd} \right).
\eea}
\begin{align}
\Theta^{\text{can}}_H = \int_H \left[T_a{}^b \left(  \ell^a (\Pi_b{}^c\delta k_c) - q_c{}^a k_b \delta \ell^c +\frac{1}{2}q^{ac}q_b{}^d \delta q_{cd} \right) - \btheta \delta \rho \right] \bm{\epsilon}_H.
\end{align}
This expression insures that the symplectic potential is covariant, i.e., it satisfies $\delta_\xi \Theta^{\mathrm{can}}_H=0$.
Let us now consider the contraction of tangential diffeomorphism $\xi = f \ell + X^A e_A$ on the canonical pre-symplectic potential. We first consider the following terms,
\begin{equation}
\lb{conjugate-to-T}
\begin{aligned}
&\ell^a (\Pi_b{}^c  \delta_\xi k_c) - (q^a{}_c  \delta_\xi \ell^c) k_b +\frac{1}{2}q^{ac}q_b{}^d \delta_\xi q_{cd}\\
&= \ell^a \Pi_b{}^c  (\xi^d \nabla_d k_c + k_d \nabla_c \xi^d) - q^a{}_c  (\xi^d \nabla_d \ell^c - \ell^d \nabla_d \xi^c) k_b  +\frac{1}{2}q^{ac}q_b{}^d (\nabla_c \xi_d +\nabla_d \xi_c) \\
& = \nabla_c \xi^d \left( \Pi_b{}^c k_d\ell^a +k_b \ell^c q_d{}^a + \frac{1}{2}q_{bd}q^{ac} + \frac{1}{2}q_b{}^c q_d{}^a  \right)  + \left( \Pi_b{}^c \ell^a \nabla_d k_c - q_c{}^a k_b \nabla_d \ell^c \right)\xi^d,
\end{aligned}
\end{equation}
where we used that $q^{ac}q_b{}^d \delta q_{cd} = q^{ac}q_b{}^d \delta g_{cd}$. Let us now consider the first term that contains $\nabla_c \xi^d$. We can show, with the help of the relation $\Pi_a{}^b = q_a{}^b +k_a \ell^b$, the following result
\begin{equation}
\begin{aligned}
\Pi_b{}^c k_d\ell^a +k_b \ell^c q_d{}^a + \frac{1}{2}q_{bd}q^{ac} + \frac{1}{2}q_b{}^c q_d{}^a & = \Pi_b{}^c (\Pi_d{}^a - q_d{}^a) +(\Pi_b{}^c - q_b{}^c) q_d{}^a + \frac{1}{2}q_{bd}q^{ac} + \frac{1}{2}q_b{}^c q_d{}^a \\
& = \Pi_b{}^c \Pi_d{}^a +  \frac{1}{2} \left( q_{bd}q^{ac} -q_b{}^c q_d{}^a \right).
\end{aligned}
\end{equation}
Note that the last term vanishes when contracting with $T_a{}^b$ by symmetry. This means that we have 
\begin{align}
T_a{}^b \nabla_c \xi^d \left( \Pi_b{}^c k_d\ell^a +k_b \ell^c q_d{}^a + \frac{1}{2}q_{bd}q^{ac} + \frac{1}{2}q_b{}^c q_d{}^a \right)  = T_a{}^b D_b \xi^a.
\end{align}
Next, the remaining term in \eqref{conjugate-to-T} that is proportional to $\xi^d$ can be written as
\begin{equation}
\begin{aligned}
\Pi_b{}^c \ell^a \nabla_d k_c - q_c{}^a k_b \nabla_d \ell^c &= \Pi_b{}^c \ell^a \nabla_d k_c - \Pi_c{}^a k_b \nabla_d \ell^c + k_c \ell^a k_b \nabla_d \ell^c \\
&= - \Pi_c{}^a k_b \nabla_d \ell^c + (\Pi_b{}^c-k_b\ell^c) \ell^a \nabla_d k_c \\
& = - \Pi_c{}^a k_b \nabla_d n^c +(2\rho \Pi_c{}^a k_b + q_{bc}\ell^a) \nabla_d k^c \\
& = - \Pi_c{}^a k_b \nabla_d n^c +(2\rho q_c{}^a k_b + q_{bc}\ell^a) \nabla_d k^c.
\end{aligned}
\end{equation}
For the first term, we recall the definition of the energy momentum tensor to write 
\begin{equation}
\begin{aligned}
- T_a{}^b k_b (\xi^d   \nabla_d n^c)\Pi_c{}^a & = - (\sW_a{}^b k_b - \sW k_a)\xi^d \sW_d{}^a \\
&= - (\omega_a \xi^d \sW_d{}^a - \sW \xi^d \omega_d) = -\xi^d T_d{}^a \omega_a.
\end{aligned}
\end{equation}
For the second term, we have that
\begin{equation}
\begin{aligned}
T_a{}^b(2\rho q_c{}^a k_b + q_{bc}\ell^a) \xi^d \nabla_d k^c &= (2\rho q_c{}^a T_a{}^b k_b + q_{bc}\ell^a T_a{}^b) \xi^dK_d{}^c \\
& = (2\rho \pi_c - \sJ_c) \xi^dK_d{}^c \\
& = \xi^c K_c{}^a(q_a{}^bD_b \rho) \\
& = \xi^a K_a{}^bD_b \rho,
\end{aligned}
\end{equation}
where we used that $\sJ_a = 2\rho \pi_a - q_a{}^b D_b \rho$ and the fact that $K_a{}^b k_b =0$. Since $\delta_\xi \rho = \xi^a D_a \rho$, we overall obtain
\begin{equation}
\begin{aligned}
I_\xi \Theta^{\text{can}}_H &= \int_H \left( T_a{}^b (D_b - \omega_b)\xi^a + \xi^a (K_a{}^b - K \Pi_a{}^b)D_b \rho  \right) \bm{\epsilon}_H \\ 
& = -\int_H \xi^a \left( D_b T_a{}^b- (K_a{}^b - K \Pi_a{}^b)D_b \rho  \right)\bm{\epsilon}_H + \bm{\rd} (\xi^b T_b{}^a \iota_a \bm{\epsilon}_H) \\
& = -\int_H G_{n\xi}\bm{\epsilon}_H +\int_{\partial H}\xi^b T_b{}^a \iota_a \bm{\epsilon}_H.\label{covxi}
\end{aligned}
\end{equation}
It is interesting to note that in this derivation we have not assumed that $D_a\rho=0$ and we have used the presence of the additional $\btheta \delta\rho$ term to prove the covariance condition \eqref{covxi}.

\section{Conclusions}

In recent years, Carrollian physics has garnered increasing attention as it emerged in a variety of situations involving null boundaries both at finite distances \cite{Penna:2018gfx,Donnay:2019jiz,Adami:2021kvx} and at asymptotic infinities \cite{Duval:2014uva,Donnay:2022aba}. The transpiration of this novel type of physics at null boundaries stems naturally from the fact that Carroll structures are universal structures of null surfaces, and the Carroll symmetry is (a part of) the symmetry group of the surfaces. In this work, we pushed this fascinating connection beyond null surfaces and argued that Carroll geometries and Carrollian physics also manifest at timelike surfaces. We demonstrated this feature for the case of the (timelike) stretched horizons located close to a finite distance null boundaries, focusing particularly on the correspondence between gravitational dynamics and hydrodynamics in the same spirit as the black hole membrane paradigm.  

Our geometrical setups relied on the rigging technique for general hypersurfaces. Let us highlight two apparent benefits of this techniques. First, by endowing on a hypersurface with a null rigged structure where a transverse vector field to the surface is null, a geometrical Carroll structure can be constructed from the elements of the rigged structure, hence providing the Carrollian picture to the intrinsic geometry of the surface, regardless of whether the surface is null or timelike. Secondly, our construction gives a unified treatment for timelike and null hypersurfaces in the way that both the stretched horizon energy-momentum tensor \eqref{e-m-def} and its conservation laws \eqref{einstein-fluid} admit non-singular limits from the timelike stretched horizon to the null boundaries. Moreover, the energy-momentum tensor \eqref{e-m-gravity}, which is interpreted as the Carrollian fluid energy-momentum tensor, allows us to establish the dictionary between gravitational degrees of freedom on the stretched horizon and Carrollian fluid quantities \eqref{dictionary}. Furthermore, the Einstein equations $\Pi_a{}^b n^cG_{ac} = 0$ is the conservation laws of the Carrollian fluid. Our result is thus a generalization of \cite{Hopfmuller:2018fni,Chandrasekaran:2021hxc,Adami:2021nnf} for the null boundaries. We have moreover shown that the correspondence between gravity and Carrollian fluids goes beyond the level of equations of motion and also manifests at the level of phase space. More precisely, the canonical part of the gravitational pre-symplectic potential \eqref{potential} decomposes the same way as the Carrollian fluid action \cite{Ciambelli:2018ojf,Freidel:2022bai}. These Carrollian hydrodynamic equations are associated with the tangential diffeomorphism  of the stretched horizon and the corresponding Noether charges take the form of the conserved charges of Carrollian fluids.  

There are of course many routes to be explored and we list here some interesting prospective research topics we think worth investigating:
\begin{enumerate}[label = \roman*), leftmargin = *]
\item {\it Thermodynamics of Carrollian fluids:} Having now established the connection between gravity and Carrollian hydrodynamics, it is then interesting to study the thermodynamics of Carrollian fluids, in turn providing a fluid route to elucidate the thermodynamical properties of the horizons. One intriguing challenge is to have a complete definition of the \emph{thermodynamical horizons}, the type of hypersurfaces that obeys all laws of (possibly non-equilibrium) thermodynamics, using the fluid analogy. Another interesting investigation is to explore the difference between Carrollian hydrodynamics and the corresponding thermodynamics of the null boundary and the stretched horizon. In the context of Carrollian fluids, the key difference between the stretched horizon and the null boundary is that the former possesses a non-zero Carrollian heat current $\sJ^A$ (see the dictionary \eqref{dictionary}) while it vanishes strictly on the latter case. It would be interesting to study how the non-zero heat current affect the thermodynamics properties of the horizon, for instance, the expression for the horizon entropy current and the entropy production. 

\item {\it Carrollian fluids at infinities:} In this work, we solely dedicated our attention to the case where the stretched horizons and the null boundary are situated at finite distances, with the example being the near-horizon geometry of black holes. It would certainly be tempting to investigate whether the similar Carrollian fluid interpretation occurs at asymptotic null infinities and, should this be the case, what the gravitational dictionary at infinity is. It is worth mentioning that there have already been a number of works that explored this null-Carroll correspondence in the context of geometry and symmetry \cite{Duval:2014uva,Duval:2014lpa,Ciambelli:2019lap}, celestial and flat holography \cite{Bagchi:2022emh,Campoleoni:2022wmf,Bagchi:2022owq, Donnay:2022aba, Donnay:2022sdg}, and Carrollian field theory \cite{Bagchi:2019xfx,Gupta:2020dtl,Bagchi:2022eav}. However, the complete Carrollian fluid picture both at the level of dynamics and the phase space has yet to be explored. 

\item {\it Stretched horizon as a radial expansion:} At null infinities, different layers of information of the null infinity phase space, symmetries, and dynamics are encoded in different orders of the radial $(1/r)$ expansion around null infinities \cite{Freidel:2021qpz, Freidel:2021dfs, Freidel:2021ytz}. This suggests that some information of a finite null boundary can be accessed by treating a stretched horizon as a radial expansion around the null boundary $(r =0)$ (also called the near-horizon expansion). One important objective is to fully derive the Einstein equations on the null boundary from the symmetry principle. To achieve that goal, we need the covariant phase space analysis of the geometry near the null boundary and the result \eqref{potential} of this current work will serve as the core basis for the near-horizon considerations. We plan to report the detailed derivations in our upcoming article \cite{forthcoming}.

\end{enumerate}

\section*{Acknowledgments}
We would like to thank C\'eline Zwikel and Luis Lehner for helpful discussions and insights. LF would also like to thank Luca Ciambelli, Niels Obers and Gerben Oling for insightful discussions on Carroll geometry.  Research at Perimeter Institute is supported in part by the Government of Canada through the Department of Innovation, Science and Economic Development Canada and by the Province of Ontario through the Ministry of Colleges and Universities. The work of LF is funded by the Natural Sciences and Engineering Research Council of Canada (NSERC) and also in part by the Alfred P. Sloan Foundation, grant FG-2020-13768. This project has received funding from the European Union's Horizon 2020 research and innovation programme under the Marie Sklodowska-Curie grant agreement No. 841923. PJ's research during his study in Canada was supported by the DPST Grant from the government of Thailand and Perimeter Institute for Theoretical Physics.

\appendix

\section{Essential elements of Carroll geometries} \lb{App:Carr}
One important result we have developed in the main text is a geometrical Carroll structure, descended from a null rigged structure, serving as a basic building block of the intrinsic geometry of the stretched horizon $H$. Here, we briefly summarize geometrical objects of the Carroll geometry that will be relevant to this present work. We will follow the notation and convention from our precursory work \cite{Freidel:2022bai} (interest readers may want to see also \cite{Ciambelli:2018xat,Ciambelli:2019lap,Petkou:2022bmz} for similar Carrollian technologies).

\subsection{Carrollian covariant derivative and curvature tensors} \lb{hor-derivative}

We have introduced the rigged covariant derivative $D_a$ as the connection on the stretched horizon $H$. There exists another layer of covariant derivative on the surface $H$ that stems from the (induced) Carroll structure of $H$. Recalling that the space $H$ has a fiber bundle structure, $p:H \to S$, and its tangent space $TH$ admits, by mean of the Ehresmann connection $\bm{k}$, the splitting into a 1-dimensional vertical subspace span by the Carrollian vector field $\ell \in \text{ker}(\bm{\rd} p)$ and the non-integrable\footnote{The non-integrability of the horizontal subspace is reflected in the commutator $[e_A,e_B] = w_{AB} \ell$ and the Frobenius theorem. It becomes however integrable when the vorticity vanishes, $w_{AB} =0$, and in that case there exists a Bondi frame.} horizontal subspace span by the basis vectors $e_A$. We then define a \emph{horizontal covariant derivative} $\sD_A$ (also called the Levi-Civita-Carroll covariant derivative \cite{Ciambelli:2018xat}) that is compatible with the sphere metric, i.e., $\sD_C q_{AB} =0$. It acts on a horizontal tensor $T = T^A{}_B e_A \otimes \bm{e}^B$ as 
\begin{align}
\sD_A T^B{}_C = e_A [T^B{}_C ]+ {}^{\sss (2)}\Gamma^B_{DA} T^D{}_C - {}^{\sss (2)}\Gamma^D_{CA} T^B{}_D,
\end{align}
and one can straightforwardly generalize it to a tensor of any degrees. The torsion-free Christoffel-Carroll symbols \cite{Ciambelli:2018xat} $\two\Gamma^A_{BC} = \two\Gamma^A_{CB} $ is defined in the same manner as the standard Christoffel symbols but instead with the sphere metric and the horizontal basis vectors,
\begin{align}
{}^{\sss (2)}\Gamma^A_{BC} := \frac{1}{2} q^{AD}\left( e_B[ q_{DC}] +e_C[ q_{BD}] - e_D [q_{BC}] \right) = q^{AD} g(e_D, \nabla_{e_B} e_C), \lb{Chris-Car}
\end{align}
where the final equality follows from $q_{AB} = g(e_A, e_B)$ and the commutator $[e_A, e_B] = w_{AB} \ell$. Let us also note that the horizontal divergence of any horizontal vector field $X = X^A e_A$ is given by the familiar formula,
\begin{align}
\sD_A X^A = \frac{1}{\sqrt{q}} e_A \left[ \sqrt{q}X^A \right].
\end{align}
The horizontal covariant derivative $\sD_A$ was defined for the timelike surface $H$ and it has a regular limit to the null boundary $N$. 

Having the horizontal covariant derivative $\sD_A$, one defines the Riemann-Carroll tensor, ${}^{\sss (2)}R^A{}_{BCD}$, whose components are determined from the commutator,
\begin{align}
[\sD_C, \sD_D] X^A = \two R^A{}_{BCD} X^B + w_{CD} \ell[X^A], \lb{Rie-Car}
\end{align}
where the last term compensates the non-integrability of the horizontal subspace. The corresponding Ricci-Carroll tensor,${}^{\sss (2)}R_{AB} :=  {}^{\sss (2)}R_{CADB} q^{CD}$, is in general not symmetric. Lastly, the Ricci-Carroll scalar is defined as ${}^{\sss (2)}R :=  {}^{\sss (2)}R_{AB} q^{AB}$.


\subsection{Volume forms and Integrations} \label{intbypart}

First, we define a volume form on the spacetime $M$ to be $\bm{\epsilon}_M = \bm{k} \wedge \bm{n} \wedge \bm{\epsilon}_S$ where $\bm{\epsilon}_S$ is the pull-back of the canonical volume form on the sphere $S$ onto the stretched horizon $H$,
\begin{align}
\bm{\epsilon}_S = \frac{1}{2}\sqrt{q} \varepsilon_{AB} \bm{e}^A \wedge \bm{e}^B = p^* \left(  \frac{1}{2}\sqrt{q} \varepsilon_{AB} \bm{\rd}\s^A \wedge \bm{\rd}\s^B \right). 
\end{align}
A volume form on the $H$ is then given by
\begin{align}
\bm{\epsilon}_H =  - \iota_k \bm{\epsilon}_M = \bm{k} \wedge \bm{\epsilon}_S, \qquad \text{and we also have that} \qquad \bm{\epsilon}_S = \iota_\ell \bm{\epsilon}_H.
\end{align}

For a function $f$ on $H$ and for a horizontal vector $X=X^A e_A$, they satisfy the following relations on the stretched horizon $H$,
\begin{align}
\bm{\rd}(f \bm{\epsilon}_S) = \left( \ell[f] + \theta f\right) \bm{\epsilon}_H, \qquad \text{and}  \qquad \bm{\rd} \left( \iota_{X} \bm{\epsilon}_H \right) = \left( \sD_A X^A +\varphi_A X^A \right) \bm{\epsilon}_H.
\end{align}
These two equations also hold on the null boundary $N$. 

In this work, we choose a boundary $\partial H$ of the stretched horizon $H$ to be located at a constant value of the coordinate $u$. This boundary is identified with the sphere $S$, meaning that $\pa H = S_u$. The Stokes theorem is therefore written as
\begin{subequations}
\label{Stokes}
\begin{align}
\int_H \left( \ell[f] + \theta f\right) \bm{\epsilon}_H &= \int_{S_u} f \bm{\epsilon}_S, \\
\int_H \left( \sD_A X^A +\varphi_A X^A \right) \bm{\epsilon}_H &= \int_{S_u} \e^\a X^A \beta_A \bm{\epsilon}_S.
\end{align}
\end{subequations}



\section{More on covariant derivatives} \lb{app:derivative}

Here, we elaborate more the relations involving the spacetime covariant derivative $\nabla_a$, the rigged covariant derivative $D_a$, and the horizontal covariant derivative $\sD_A$. First, let us provide the general form of the spacetime covariant derivative of the tangential vector $\ell$, the transverse vector $k$, and their combination $n = \ell + 2\rho k$ that will become handy in the computations,
\begin{align}
\nabla_a \ell^b = \ & \theta_a{}^b + (\pi_a + \kappa k_a) \ell^b - k_a(\sJ^b-2 (\pi^b+\varphi^b)\rho )  + 
\left( \sJ_a   -k_a (\ell-2\kappa)[\rho] \right)k^b 
\nonumber \\
& - n_a\left( (k[\rho] + 2\rho \bar\kappa) k^b + (\pi^b + \varphi^b) +  \bar{\kappa} \ell^b\right), \lb{nabla-l}\\
\nabla_a k^b = \ & \btheta_a{}^b -(\pi_a + \kappa k_a)k^b - k_a (\pi^b + \varphi^b)  + \bar{\kappa} n_a k^b.  \lb{nabla-k}
\\
\nabla_a n^b = \ & (\theta_a{}^b+2\rho\bar\theta_a{}^b) + 
(\pi_a + \kappa k_a) \ell^b 
- k_a \sJ^b  +D_a\rho k^b
\nonumber \\
&
- n_a\left( - k[\rho] k^b + (\pi^b + \varphi^b) +  \bar{\kappa} \ell^b\right).
\end{align}
We emphasize here again that the Carrollian current is given in general by \eqref{general-J}, $\sJ_a =-(\sD_a -2\pi_a)[\rho]$. The divergence of these vectors are
\begin{align}
\nabla_a \ell^a = \theta + \kappa - (k[\rho]+ 2\rho \bar\kappa), \qquad \text{and} \qquad \nabla_a k^a = \btheta + \bar{\kappa}. 
\end{align}
The projections of \eqref{nabla-l} and \eqref{nabla-k} are thus given by 
\begin{align}
D_a \ell^b & := \Pi_a{}^c \Pi_d{}^b \nabla_c \ell^d =  \theta_a{}^b + \pi_a \ell^b + k_aA^b + \kappa k_a \ell^b, \lb{D-ell} \\
K_a{}^b & := \Pi_a{}^c \Pi_d{}^b \nabla_c k^d = \btheta_a{}^b - k_a (\pi^b + \varphi^b), \lb{D-k}
\end{align}
where we recalled the acceleration $A^a = (\sD^a + 2\varphi^a)\rho$. Their traces are
\begin{align}
D_a \ell^a = \theta + \kappa, \qquad \text{and} \qquad D_a k^a = \btheta. 
\end{align}

As we have seen, there are three layers of covariant derivatives, $\nabla_a, D_a$ and $\sD_a$. To connect them, we first look at the spacetime covariant derivative of the horizontal basis $e_A$ along another horizontal basis. One can verify that it is given by
\begin{align}
\nabla_{e_A} e_B = \two\Gamma^C_{AB} e_C - \btheta_{AB} \ell - \theta_{AB} k. \lb{del-e}
\end{align}
Using the decomposition of the spacetime metric \eqref{4metric} and the Leibniz rule, we express the spacetime divergence of the horizontal basis as
\begin{equation}
\begin{aligned}
\nabla_a e_A{}^a & = \left(n_a k^b + k_a \ell^b + q^{BC} e_{Ba} e_C{}^b    \right) \nabla_b e_A{}^a 
 = \two\Gamma^B_{BA}  + 2(\varphi_A + \pi_A). 
\end{aligned}
\end{equation}
Observe that if we set the scale factor $\bar{\alpha} =0$, we simply have that $2(\varphi_A + \pi_A) = \varphi_A$. Following from these results, the covariant derivative of a generic horizontal vector fields $X^a := X^A e_A{}^a$ projected onto the horizontal subspace is 
\begin{align}
e^B{}_a\nabla_{e_A} X^a & = e_A[ X^B] + X^C e^B{}_b \nabla_{e_A} e_C{}^b = \sD_A X^B. 
\end{align}
Furthermore, the spacetime divergence of the horizontal vector is 
\begin{equation}
\begin{aligned}
\nabla_a \left( X^A e_A{}^a \right) &= e_A[ X^A] + X^A \nabla_a e_A{}^a = \left(\sD_A +2(\pi_A +\varphi_A) \right)X^A.
\end{aligned}
\end{equation} 

In addition, let us also look at the rigged covariant derivative of the horizontal basis. We can show by recalling that $\Pi_a{}^b = q_a{}^b + k_a \ell^b$ and $q_a{}^b = e^A{}_a e_A{}^b$ the following relation
\begin{equation}
\lb{D-e_A}
\begin{aligned}
D_b e_A{}^a &=\Pi_b{}^d \Pi_c{}^a \nabla_d e_A{}^c \\
&= (q_b{}^d + k_b\ell^d) \nabla_d e_A{}^c (q_c{}^a + k_c\ell^a) \\
& = q^{CD}g(e_D, \nabla_{e_B} e_A) e^B{}_b e_C{}^a + g(k,\nabla_{e_B} e_A) e^B{}_b \ell^a + q^{BC}g(e_C, \nabla_{\ell} e_A) k_b e_B{}^a + g(k,\nabla_\ell e_A) k_b \ell^a \\
&= \two\Gamma^C_{BA} e_C{}^a e^B{}_b +(
- \btheta_{BA}  e^B{}_b + (\pi_A + \varphi_A) k_b )\ell^a + \theta_A{}^B e_B{}^a k_b. 
\end{aligned} 
\end{equation}
The rigged divergence of the horizontal basis is simply the trace, 
\begin{align}
D_a e_A{}^a = \two\Gamma^B_{BA} + (\pi_A+\varphi_A).
\end{align}
With this, the rigged divergence of a horizontal vector field $X^a = X^A e_A{}^a$ is then
\begin{align}
D_a X^a = D_a (X^A e_A{}^a) = \sD_A X^A + (\pi_A + \varphi_A) X^A.
\end{align}

For completeness, let us also compute the rigged covariant derivative of the co-frame $\bm{e}^A$. Using that $D_a (\ell^b e^A{}_b) = \ell^b D_a e^A{}_b + e^A{}_b D_a \ell^b =0$ and $D_a (e_B{}^b e^A{}_b) = e_B{}^b D_a e^A{}_b + e^A{}_b D_a e_B{}^b =0$ we hence write 
\begin{align}
D_a e^A{}_b &= - (e^A{}_c D_a \ell^c)k_b - (e^A{}_b D_a e_B{}^b) e^B{}_b \\
& = - (\theta_a{}^A + k_a A^A)k_b - \two\Gamma^A_{ab}-\theta_b{}^A k_a.
\end{align}
Following from $q_c{}^b = e^A{}_c e_A{}^b$, we can then show that the rigged covariant derivative of the null Carrollian metric is 
\begin{align}
D_a q_c{}^b &= e^A{}_c D_ae_A{}^b + e_A{}^b D_a e^A{}_c \\
&= \left( \two\Gamma^b_{ac} +(- \btheta_{ac}+(\pi_a + \varphi_c)k_a)\ell^b + k_a \theta_c{}^b \right) - \left( (\theta_a{}^b + k_a A^b)k_c + \two\Gamma^b_{ac}+\theta_c{}^b k_a\right) \\
& = (- \btheta_{ac}+(\pi_a + \varphi_c)k_a)\ell^b - (\theta_a{}^b + k_a A^b)k_c. 
\end{align}
This result can also be obtain by simply using that $q_c{}^b = \Pi_c{}^b - k_c \ell^b$ and that $D_a \Pi_c{}^b =0$. 

\section{Derivation of the pre-symplectic potential} \lb{potential-derive}

In this section, we present in detail how to write the gravitational pre-symplectic potential in terms of Carrollian fluid variables. For the Einstein-Hilbert gravity, the pre-symplectic potential evaluated on the stretched horizon $H$ is given by
\begin{equation}
\lb{EH-potential}
\begin{aligned}
\bm{\Theta}_H = -\Theta^a n_a \bm{\epsilon}_H, \qquad \text{where} \qquad  \Theta^a = \frac{1}{2} \left( g^{ac} \nabla^b \delta g_{bc} - \nabla^a (g^{bc} \delta g_{bc}) \right),
\end{aligned}
\end{equation}
and we recalled that $\bm{\epsilon}_H := -\iota_k \bm{\epsilon}_M$. To evaluate the pre-symplectic potential, one starts with the variation of the spacetime metric which, by using the decomposition \eqref{4metric}, can be expressed as follows,  
\begin{equation}
\lb{delta-g}
\begin{aligned}
\delta g_{ab} &= \delta q_{ab} + 2\delta n_{(a} k_{b)} + 2n_{(a} \delta k_{b)} -4\rho k_{(a} \delta k_{b)} - 2 \delta \rho k_a k_b \\
&= \delta q_{ab}   + 2k_{(a} \delta n_{b)} + 2\ell_{(a} \delta k_{b)} - 2\delta \rho k_a k_b,
\end{aligned}
\end{equation}
where we recalled that $\ell_a = n_a - 2\rho k_a$. The trace of the metric variation is then
\begin{align}
g^{bc}\delta g_{bc} = 2(\delta \alpha + \delta \bar{\alpha} + \delta \ln \sqrt{q}) = 2(\bbdelta \alpha + \delta \bar{\alpha} + \bbdelta \ln \sqrt{q}).
\end{align}

The task now is to consider the first term, which is $n^a \nabla^b \delta g_{ab}$, in the gravitational pre-symplectic potential. Let us evaluate each term in \eqref{delta-g} separately as follows:
\begin{itemize}[label = $\square$, leftmargin = *]
\item First, using that $\delta n_a = \delta \bar{\alpha} n_a$ and the Leibniz rule, we can show that 
\begin{equation}
\begin{aligned}
n^a\nabla^b \left(2k_{(a} \delta n_{b)} \right) &= n^a \nabla^b \left( \delta \bar{\alpha}(k_a n_b + n_a k_b ) \right)  \\
& = \nabla_a \left(  (n^a + 2\rho k^a) \delta \bar{\alpha}\right) -(n_a\nabla_k n^a + k_a\nabla_n n^a) \delta \bar{\alpha} \\
& = \left( n + 2\rho k + \nabla_a n^a + 2 \rho \nabla_a k^a + 2 k[\rho] \right)[\delta \bar{\alpha}] - \left( k[\rho] + \kappa - 2\rho \bar{\kappa}\right) \delta \bar{\alpha} \\
& = \left(\ell + 4\rho k+ \theta + 4\rho(\btheta + \bar{\kappa})+ 2 k[\rho]  \right)[\delta \bar{\alpha}],
\end{aligned}
\end{equation}
where we used the formulae for the divergence $\nabla_a n^a = \theta + 2\rho \btheta + \kappa + k[\rho]$ and $\nabla_a k^a = \btheta + \bar{\kappa}$. 

\item Using the variation of the Ehresmann connection $\delta k_a = \bbdelta \alpha k_a - \e^\alpha \bbdelta \beta_A e^A{}_a$, we show the folllowing
\begin{equation}
\begin{aligned}
n^a\nabla^b \left(2\ell_{(a} \delta k_{b)} \right) 
&= \nabla_a \left( (n^b \delta k_b) \ell^a\right) - \ell_a \delta k _b (\nabla^a n^b + \nabla^b n^a) \\
&= \left(\ell + \nabla_a \ell^a \right)\left[\bbdelta \alpha\right]  - (k_a \nabla_\ell n^a + \ell_a\nabla_k n^a) \bbdelta \alpha + \e^\a \left( e^A{}_a \nabla_\ell n^a + q^{AB} \ell_a \nabla_{e_B} n^a \right) \bbdelta \beta_A \\
& = \left(\ell + \theta +\kappa - k[\rho] - 2\rho \bar{\kappa}\right)[\bbdelta \alpha ] - \left(\kappa+ k[\rho] +2\rho\bar{\kappa} \right) \bbdelta \alpha -  \e^\a \left(\sJ^A - (\sD^A - 2 \pi^A)[\rho] \right)\bbdelta \beta_A \\
& = \left(\ell + \theta -2( k[\rho] +2\rho \bar{\kappa})\right)[\bbdelta \alpha ] -   2\e^\a \sJ^A\bbdelta \beta_A \\
& = \left(\ell + \theta -2( \kappa - \ell[\bar{\alpha}])\right)[\bbdelta \alpha ] -   2\e^\a \sJ^A\bbdelta \beta_A,
\end{aligned}
\end{equation}
where we used the relation $\sJ_A = - (\sD_A -2\pi_A)[\rho]$. 

\item Using the Leibniz rule, we have that 
\begin{equation}
\begin{aligned}
-2n^a\nabla^b \left(\delta \rho k_a k_b \right) 
&= -2\nabla_a \left( \delta \rho k^a\right) +2\left( k_a \nabla_k n^a \right) \delta \rho \\
& = -2 \left( k + \nabla_a k^a \right)[\delta \rho] +2\left( k_a \nabla_k n^a \right) \delta \rho \\
& = -2 \left( k + \btheta+2\bar{\kappa} \right)[\delta \rho] 
\end{aligned}
\end{equation}

\item Lastly, the term involving variation of the null metric can be evaluate as follows,
\begin{equation}
\begin{aligned}
n^a\nabla^b \delta q_{ab} &= \nabla^a \left( n^b \delta q_{ab}\right) - (\nabla^a n^b)\delta q_{ab} \\
&= -\nabla_a \left(\e^{-\a} \bbdelta V^A e_A{}^a \right) + \e^{-\alpha}\left(e_{Aa} \nabla_k n^a +k_a\nabla_{e_A} n^a\right) \bbdelta V^A - \left(q^{BC}e^A{}_a \nabla_{e_C} n^a \right)\bbdelta q_{AB} \\
& = - \left( \sD_A +2\pi_A+ 2\varphi_A \right)\left(\e^{-\a} \bbdelta V^A\right) -\e^{-\a}\varphi_A  \bbdelta V^A-\left( \sN^{AB} + \frac{1}{2}\sE q^{AB} \right)\bbdelta q_{AB}\\
& = - \left( \sD_A +\varphi_A \right)\left(\e^{-\a} \bbdelta V^A\right) +2\e^{-\a}\left(\pi_A -e_A[\bar{\alpha}]  \right)\bbdelta V^A-\left( \sN^{AB} + \frac{1}{2}\sE q^{AB} \right)\bbdelta q_{AB},
\end{aligned}
\end{equation}
where to obtain the last equality, we used the dictionary \eqref{dictionary} that $e_A[\bar{\alpha}] = 2\pi_A + \varphi_A$. 
\end{itemize}
The second term in the pre-symplectic potential \eqref{EH-potential} is simply the derivative of the trace of the metric variation along the direction of the vector $n^a$ that can be expressed as 
\begin{equation}
\begin{aligned}
n^a \nabla_a \left(g^{bc}\delta g_{bc}\right) &= 2n[\bbdelta \alpha + \delta \bar{\alpha} + \bbdelta \ln \sqrt{q}] \\
&= 2\ell[\bbdelta \alpha + \delta \bar{\alpha} + \bbdelta \ln \sqrt{q}] + 4\rho k[\delta \alpha + \delta \bar{\alpha} + \delta \ln \sqrt{q} ] \\
&= 2\ell[\bbdelta \alpha + \delta \bar{\alpha} + \bbdelta \ln \sqrt{q}] + 4\rho k[\delta \alpha + \delta \bar{\alpha} ] + 4\rho \left( \delta \btheta + \btheta \delta \bar{\alpha} \right),
\end{aligned}
\end{equation}
where we used the Leibniz rule to write $k[\delta \ln \sqrt{q}] = \delta \left(  k[\ln \sqrt{q}]  \right) - \delta k [ \ln \sqrt{q}]$ and that $\delta k = -\delta \bar{\alpha} k$ and $\btheta = k[ \ln \sqrt{q}]$. 

Collecting all the results, we arrive at the following expression for the pre-symplectic potential \eqref{EH-potential}
\begin{equation}
\begin{aligned}
2\Theta^a n_a  = \ & 2(\theta - \kappa) \bbdelta \alpha - 2 \e^\a \sJ^A \bbdelta \beta_A +2\e^{-\a}\pi_A  \bbdelta V^A -\left( \sN^{AB} + \frac{1}{2}(\sE-2\theta) q^{AB} \right)\bbdelta q_{AB} - 2 \btheta \delta \rho - 4\rho \delta \btheta \\
& - 2\left( \ell[\delta \bar{\alpha}] - \ell[\bar{\alpha}]\bbdelta \alpha + \e^{-\a}\bbdelta V^A e_A[\bar{\alpha}] + k [\delta \rho]  - k[\rho] \delta \bar{\alpha} +  2\bar{\kappa}\delta \rho +2\rho \left( k[\delta \alpha] - \bar{\kappa} \delta \bar{\alpha}   \right) \right) \\
& - \left( \sD_A + \varphi_A \right)\left(\e^{-\a} \bbdelta V^A\right) + \left( \ell + \theta \right)[\delta \bar{\alpha} - \bbdelta \alpha - 2 \bbdelta \ln \sqrt{q}].
\end{aligned}
\end{equation}
The term on the second line is actually the variation of the surface gravity $\kappa$, which one can check straightforwardly, by recalling the expression $\kappa = \ell[\bar{\alpha}] + k[\rho] + 2\rho \bar{\kappa}$ and $\bar{\kappa} = k[\alpha]$, that 
\begin{align}
\delta \kappa &= \ell[\delta \bar{\alpha}] + \delta\ell[\bar{\alpha}]+ k [\delta \rho]  + \delta k[\rho]  +2 \bar{\kappa} \delta \rho + 2\rho \delta \bar{\kappa}  \\
&= \ell[\delta \bar{\alpha}]   - \ell[\bar{\alpha}]\bbdelta \alpha + \e^{-\a}\bbdelta V^A e_A[\bar{\alpha}] + k [\delta \rho]  - k[\rho] \delta \bar{\alpha} +2 \bar{\kappa} \delta \rho + 2\rho \delta \bar{\kappa} \\
\delta \bar{\kappa} &= k[\delta \alpha] - \bar{\kappa} \delta \bar{\alpha}. 
\end{align}
Then, using the Leibniz rule and that $\delta \bm{\epsilon}_H = \left( \bbdelta \alpha + \bbdelta \ln \sqrt{q} \right) \bm{\epsilon}_H$, we can finally show that 
\begin{equation}
\begin{aligned}
\bm{\Theta}_H  = \ & \left(-\sE \bbdelta \alpha + 2 \e^\a \sJ^A \bbdelta \beta_A -2\e^{-\a}\pi_A  \bbdelta V^A +\frac{1}{2}\left( \sN^{AB} +\sP q^{AB} \right)\bbdelta q_{AB} - \btheta \delta \rho \right)\bm{\epsilon}_H \\
& + \delta \left((\kappa + 2\rho \btheta) \bm{\epsilon}_H\right) + \left( \frac{1}{2}(\delta \alpha -\delta \bar{\alpha})  +  \delta \ln \sqrt{q}\right)\bm{\epsilon}_S.
\end{aligned}
\end{equation}
where we used that $\sE = \theta + 2\rho \btheta$ and $\sP = -\kappa -\frac{1}{2}\sE$. Finally, using that $(\delta \ln \sqrt{q}) \bm{\epsilon}_S = \delta \bm{\epsilon}_S = \delta (\theta \bm{\epsilon}_H)$, we obtain
\begin{equation}
\begin{aligned}
\bm{\Theta}_H  = \ & \left(-\sE \bbdelta \alpha + 2 \e^\a \sJ^A \bbdelta \beta_A -2\e^{-\a}\pi_A  \bbdelta V^A +\frac{1}{2}\left( \sN^{AB} +\sP q^{AB} \right)\bbdelta q_{AB} - \btheta \delta \rho \right)\bm{\epsilon}_H \\
& + \delta \left((\kappa + \sE) \bm{\epsilon}_H\right) +  \frac{1}{2}(\delta \alpha -\delta \bar{\alpha}) \bm{\epsilon}_S.
\end{aligned}
\end{equation}

\bibliography{Biblio.bib}

\providecommand{\href}[2]{#2}\begingroup\raggedright\begin{thebibliography}{10}

\bibitem{Maldacena:1997re}
J.~M. Maldacena, \emph{{The Large N limit of superconformal field theories and
  supergravity}}, \href{http://dx.doi.org/10.1023/A:1026654312961}{\emph{Adv.
  Theor. Math. Phys.} {\bfseries 2} (1998) 231--252},
  [\href{https://arxiv.org/abs/hep-th/9711200}{{\ttfamily hep-th/9711200}}].

\bibitem{Witten:1998qj}
E.~Witten, \emph{{Anti-de Sitter space and holography}},
  \href{http://dx.doi.org/10.4310/ATMP.1998.v2.n2.a2}{\emph{Adv. Theor. Math.
  Phys.} {\bfseries 2} (1998) 253--291},
  [\href{https://arxiv.org/abs/hep-th/9802150}{{\ttfamily hep-th/9802150}}].

\bibitem{Strominger:2017zoo}
A.~Strominger, \emph{{Lectures on the Infrared Structure of Gravity and Gauge
  Theory}},  [\href{https://arxiv.org/abs/1703.05448}{{\ttfamily 1703.05448}}].

\bibitem{Raclariu:2021zjz}
A.-M. Raclariu, \emph{{Lectures on Celestial Holography}},
  [\href{https://arxiv.org/abs/2107.02075}{{\ttfamily 2107.02075}}].

\bibitem{Pasterski:2021rjz}
S.~Pasterski, \emph{{Lectures on Celestial Amplitudes}},
  \href{http://dx.doi.org/10.1140/epjc/s10052-021-09846-7}{\emph{Eur. Phys. J.
  C} {\bfseries 81} (2021) 1062},
  [\href{https://arxiv.org/abs/2108.04801}{{\ttfamily 2108.04801}}].

\bibitem{Donnelly:2016auv}
W.~Donnelly and L.~Freidel, \emph{{Local Subsystems in Gauge Theory and
  Gravity}}, \href{http://dx.doi.org/10.1007/JHEP09(2016)102}{\emph{JHEP}
  {\bfseries 09} (2016) 102},
  [\href{https://arxiv.org/abs/1601.04744}{{\ttfamily 1601.04744}}].

\bibitem{Speranza:2017gxd}
A.~J. Speranza, \emph{{Local phase space and edge modes for
  diffeomorphism-invariant theories}},
  \href{http://dx.doi.org/10.1007/JHEP02(2018)021}{\emph{JHEP} {\bfseries 02}
  (2018) 021}, [\href{https://arxiv.org/abs/1706.05061}{{\ttfamily
  1706.05061}}].

\bibitem{Geiller:2017xad}
M.~Geiller, \emph{{Edge modes and corner ambiguities in 3d
  Chern\textendash{}Simons theory and gravity}},
  \href{http://dx.doi.org/10.1016/j.nuclphysb.2017.09.010}{\emph{Nucl. Phys. B}
  {\bfseries 924} (2017) 312--365},
  [\href{https://arxiv.org/abs/1703.04748}{{\ttfamily 1703.04748}}].

\bibitem{Geiller:2017whh}
M.~Geiller, \emph{{Lorentz-Diffeomorphism Edge Modes in 3D Gravity}},
  \href{http://dx.doi.org/10.1007/JHEP02(2018)029}{\emph{JHEP} {\bfseries 02}
  (2018) 029}, [\href{https://arxiv.org/abs/1712.05269}{{\ttfamily
  1712.05269}}].

\bibitem{Freidel:2020xyx}
L.~Freidel, M.~Geiller and D.~Pranzetti, \emph{{Edge Modes of Gravity. Part I.
  Corner Potentials and Charges}},
  \href{http://dx.doi.org/10.1007/JHEP11(2020)026}{\emph{JHEP} {\bfseries 11}
  (2020) 026}, [\href{https://arxiv.org/abs/2006.12527}{{\ttfamily
  2006.12527}}].

\bibitem{Freidel:2020svx}
L.~Freidel, M.~Geiller and D.~Pranzetti, \emph{{Edge Modes of Gravity. Part II.
  Corner Metric and Lorentz Charges}},
  \href{http://dx.doi.org/10.1007/JHEP11(2020)027}{\emph{JHEP} {\bfseries 11}
  (2020) 027}, [\href{https://arxiv.org/abs/2007.03563}{{\ttfamily
  2007.03563}}].

\bibitem{Freidel:2020ayo}
L.~Freidel, M.~Geiller and D.~Pranzetti, \emph{{Edge modes of gravity. Part
  III. Corner simplicity constraints}},
  \href{http://dx.doi.org/10.1007/JHEP01(2021)100}{\emph{JHEP} {\bfseries 01}
  (2021) 100}, [\href{https://arxiv.org/abs/2007.12635}{{\ttfamily
  2007.12635}}].

\bibitem{Ciambelli:2021vnn}
L.~Ciambelli and R.~G. Leigh, \emph{{Isolated surfaces and symmetries of
  gravity}}, \href{http://dx.doi.org/10.1103/PhysRevD.104.046005}{\emph{Phys.
  Rev. D} {\bfseries 104} (2021) 046005},
  [\href{https://arxiv.org/abs/2104.07643}{{\ttfamily 2104.07643}}].

\bibitem{Donnelly:2020xgu}
W.~Donnelly, L.~Freidel, S.~F. Moosavian and A.~J. Speranza,
  \emph{{Gravitational Edge Modes, Coadjoint Orbits, and Hydrodynamics}},
  \href{http://dx.doi.org/10.1007/JHEP09(2021)008}{\emph{JHEP} {\bfseries 09}
  (2021) 008}, [\href{https://arxiv.org/abs/2012.10367}{{\ttfamily
  2012.10367}}].

\bibitem{Balachandran:1994up}
A.~P. Balachandran, L.~Chandar and A.~Momen, \emph{{Edge States in Gravity and
  Black Hole Physics}},
  \href{http://dx.doi.org/10.1016/0550-3213(95)00622-2}{\emph{Nucl. Phys. B}
  {\bfseries 461} (1996) 581--596},
  [\href{https://arxiv.org/abs/gr-qc/9412019}{{\ttfamily gr-qc/9412019}}].

\bibitem{Geiller:2019bti}
M.~Geiller and P.~Jai-akson, \emph{{Extended actions, dynamics of edge modes,
  and entanglement entropy}},
  \href{http://dx.doi.org/10.1007/JHEP09(2020)134}{\emph{JHEP} {\bfseries 09}
  (2020) 134}, [\href{https://arxiv.org/abs/1912.06025}{{\ttfamily
  1912.06025}}].

\bibitem{Freidel:2021cjp}
L.~Freidel, R.~Oliveri, D.~Pranzetti and S.~Speziale, \emph{{Extended corner
  symmetry, charge bracket and Einstein\textquoteright{}s equations}},
  \href{http://dx.doi.org/10.1007/JHEP09(2021)083}{\emph{JHEP} {\bfseries 09}
  (2021) 083}, [\href{https://arxiv.org/abs/2104.12881}{{\ttfamily
  2104.12881}}].

\bibitem{Compere:2019bua}
G.~Comp\`ere, A.~Fiorucci and R.~Ruzziconi, \emph{{The $\Lambda$-BMS$_4$ group
  of dS$_4$ and new boundary conditions for AdS$_4$}},
  \href{http://dx.doi.org/10.1088/1361-6382/ab3d4b}{\emph{Class. Quant. Grav.}
  {\bfseries 36} (2019) 195017},
  [\href{https://arxiv.org/abs/1905.00971}{{\ttfamily 1905.00971}}].

\bibitem{Fernandez-Alvarez:2021yog}
F.~Fern\'andez-\'Alvarez and J.~M.~M. Senovilla, \emph{{Asymptotic structure
  with a positive cosmological constant}},
  \href{http://dx.doi.org/10.1088/1361-6382/ac395b}{\emph{Class. Quant. Grav.}
  {\bfseries 39} (2022) 165012},
  [\href{https://arxiv.org/abs/2105.09167}{{\ttfamily 2105.09167}}].

\bibitem{Geiller:2022vto}
M.~Geiller and C.~Zwikel, \emph{{The partial Bondi gauge: Further enlarging the
  asymptotic structure of gravity}},
  [\href{https://arxiv.org/abs/2205.11401}{{\ttfamily 2205.11401}}].

\bibitem{Damour:1978cg}
T.~Damour, \emph{{Black Hole Eddy Currents}},
  \href{http://dx.doi.org/10.1103/PhysRevD.18.3598}{\emph{Phys. Rev. D}
  {\bfseries 18} (1978) 3598--3604}.

\bibitem{Thorne:1986iy}
K.~S. Thorne, R.~H. Price and D.~A. Macdonald, eds., \emph{{Black Holes: the
  Membrane Paradigm}}.
\newblock 1986.

\bibitem{Price:1986yy}
R.~H. Price and K.~S. Thorne, \emph{{Membrane Viewpoint on Black Holes:
  Properties and Evolution of the Stretched Horizon}},
  \href{http://dx.doi.org/10.1103/PhysRevD.33.915}{\emph{Phys. Rev. D}
  {\bfseries 33} (1986) 915--941}.

\bibitem{Anninos:1993zj}
P.~Anninos, D.~Hobill, E.~Seidel, L.~Smarr and W.-M. Suen, \emph{{The Collision
  of Two Black Holes}},
  \href{http://dx.doi.org/10.1103/PhysRevLett.71.2851}{\emph{Phys. Rev. Lett.}
  {\bfseries 71} (1993) 2851--2854},
  [\href{https://arxiv.org/abs/gr-qc/9309016}{{\ttfamily gr-qc/9309016}}].

\bibitem{Anninos:1994gp}
P.~Anninos, D.~Hobill, E.~Seidel, L.~Smarr and W.-M. Suen, \emph{{The Headon
  Collision of Two Equal Mass Black Holes}},
  \href{http://dx.doi.org/10.1103/PhysRevD.52.2044}{\emph{Phys. Rev. D}
  {\bfseries 52} (1995) 2044--2058},
  [\href{https://arxiv.org/abs/gr-qc/9408041}{{\ttfamily gr-qc/9408041}}].

\bibitem{Faulkner:2010jy}
T.~Faulkner, H.~Liu and M.~Rangamani, \emph{{Integrating Out Geometry:
  Holographic Wilsonian RG and the Membrane Paradigm}},
  \href{http://dx.doi.org/10.1007/JHEP08(2011)051}{\emph{JHEP} {\bfseries 08}
  (2011) 051}, [\href{https://arxiv.org/abs/1010.4036}{{\ttfamily 1010.4036}}].

\bibitem{Bredberg:2011jq}
I.~Bredberg, C.~Keeler, V.~Lysov and A.~Strominger, \emph{{From Navier-Stokes
  to Einstein}}, \href{http://dx.doi.org/10.1007/JHEP07(2012)146}{\emph{JHEP}
  {\bfseries 07} (2012) 146},
  [\href{https://arxiv.org/abs/1101.2451}{{\ttfamily 1101.2451}}].

\bibitem{Bhattacharyya:2015dva}
S.~Bhattacharyya, A.~De, S.~Minwalla, R.~Mohan and A.~Saha, \emph{{A Membrane
  Paradigm at Large D}},
  \href{http://dx.doi.org/10.1007/JHEP04(2016)076}{\emph{JHEP} {\bfseries 04}
  (2016) 076}, [\href{https://arxiv.org/abs/1504.06613}{{\ttfamily
  1504.06613}}].

\bibitem{Bhattacharyya:2007vjd}
S.~Bhattacharyya, V.~E. Hubeny, S.~Minwalla and M.~Rangamani, \emph{{Nonlinear
  Fluid Dynamics from Gravity}},
  \href{http://dx.doi.org/10.1088/1126-6708/2008/02/045}{\emph{JHEP} {\bfseries
  02} (2008) 045}, [\href{https://arxiv.org/abs/0712.2456}{{\ttfamily
  0712.2456}}].

\bibitem{Son:2007vk}
D.~T. Son and A.~O. Starinets, \emph{{Viscosity, Black Holes, and Quantum Field
  Theory}},
  \href{http://dx.doi.org/10.1146/annurev.nucl.57.090506.123120}{\emph{Ann.
  Rev. Nucl. Part. Sci.} {\bfseries 57} (2007) 95--118},
  [\href{https://arxiv.org/abs/0704.0240}{{\ttfamily 0704.0240}}].

\bibitem{Rangamani:2009xk}
M.~Rangamani, \emph{{Gravity and Hydrodynamics: Lectures on the fluid-gravity
  correspondence}},
  \href{http://dx.doi.org/10.1088/0264-9381/26/22/224003}{\emph{Class. Quant.
  Grav.} {\bfseries 26} (2009) 224003},
  [\href{https://arxiv.org/abs/0905.4352}{{\ttfamily 0905.4352}}].

\bibitem{Hubeny:2010wp}
V.~E. Hubeny, \emph{{The Fluid/Gravity Correspondence: a new perspective on the
  Membrane Paradigm}},
  \href{http://dx.doi.org/10.1088/0264-9381/28/11/114007}{\emph{Class. Quant.
  Grav.} {\bfseries 28} (2011) 114007},
  [\href{https://arxiv.org/abs/1011.4948}{{\ttfamily 1011.4948}}].

\bibitem{Hubeny:2011hd}
V.~E. Hubeny, S.~Minwalla and M.~Rangamani, \emph{{The fluid/gravity
  correspondence}},  in \emph{{Theoretical Advanced Study Institute in
  Elementary Particle Physics}: {String theory and its Applications: From meV
  to the Planck Scale}}, pp.~348--383, 2012.
\newblock \href{https://arxiv.org/abs/1107.5780}{{\ttfamily 1107.5780}}.

\bibitem{Iqbal:2008by}
N.~Iqbal and H.~Liu, \emph{{Universality of the Hydrodynamic Limit in AdS/CFT
  and the Membrane Paradigm}},
  \href{http://dx.doi.org/10.1103/PhysRevD.79.025023}{\emph{Phys. Rev. D}
  {\bfseries 79} (2009) 025023},
  [\href{https://arxiv.org/abs/0809.3808}{{\ttfamily 0809.3808}}].

\bibitem{Eling:2009pb}
C.~Eling, I.~Fouxon and Y.~Oz, \emph{{The Incompressible Navier-Stokes
  Equations from Membrane Dynamics}},
  \href{http://dx.doi.org/10.1016/j.physletb.2009.09.028}{\emph{Phys. Lett. B}
  {\bfseries 680} (2009) 496--499},
  [\href{https://arxiv.org/abs/0905.3638}{{\ttfamily 0905.3638}}].

\bibitem{Nickel:2010pr}
D.~Nickel and D.~T. Son, \emph{{Deconstructing Holographic Liquids}},
  \href{http://dx.doi.org/10.1088/1367-2630/13/7/075010}{\emph{New J. Phys.}
  {\bfseries 13} (2011) 075010},
  [\href{https://arxiv.org/abs/1009.3094}{{\ttfamily 1009.3094}}].

\bibitem{Gregory:1993vy}
R.~Gregory and R.~Laflamme, \emph{{Black strings and p-branes are unstable}},
  \href{http://dx.doi.org/10.1103/PhysRevLett.70.2837}{\emph{Phys. Rev. Lett.}
  {\bfseries 70} (1993) 2837--2840},
  [\href{https://arxiv.org/abs/hep-th/9301052}{{\ttfamily hep-th/9301052}}].

\bibitem{Cardoso:2006ks}
V.~Cardoso and O.~J.~C. Dias, \emph{{Rayleigh-Plateau and Gregory-Laflamme
  instabilities of black strings}},
  \href{http://dx.doi.org/10.1103/PhysRevLett.96.181601}{\emph{Phys. Rev.
  Lett.} {\bfseries 96} (2006) 181601},
  [\href{https://arxiv.org/abs/hep-th/0602017}{{\ttfamily hep-th/0602017}}].

\bibitem{Freidel:2014qya}
L.~Freidel and Y.~Yokokura, \emph{{Non-equilibrium thermodynamics of
  gravitational screens}},
  \href{http://dx.doi.org/10.1088/0264-9381/32/21/215002}{\emph{Class. Quant.
  Grav.} {\bfseries 32} (2015) 215002},
  [\href{https://arxiv.org/abs/1405.4881}{{\ttfamily 1405.4881}}].

\bibitem{Unruh:1980cg}
W.~G. Unruh, \emph{{Experimental black hole evaporation}},
  \href{http://dx.doi.org/10.1103/PhysRevLett.46.1351}{\emph{Phys. Rev. Lett.}
  {\bfseries 46} (1981) 1351--1353}.

\bibitem{Freidel:2021qpz}
L.~Freidel and D.~Pranzetti, \emph{{Gravity from Symmetry: Duality and
  Impulsive Waves}},
  \href{http://dx.doi.org/10.1007/JHEP04(2022)125}{\emph{JHEP} {\bfseries 04}
  (2022) 125}, [\href{https://arxiv.org/abs/2109.06342}{{\ttfamily
  2109.06342}}].

\bibitem{Freidel:2021dfs}
L.~Freidel, D.~Pranzetti and A.-M. Raclariu, \emph{{Sub-subleading soft
  graviton theorem from asymptotic Einstein\textquoteright{}s equations}},
  \href{http://dx.doi.org/10.1007/JHEP05(2022)186}{\emph{JHEP} {\bfseries 05}
  (2022) 186}, [\href{https://arxiv.org/abs/2111.15607}{{\ttfamily
  2111.15607}}].

\bibitem{Freidel:2021ytz}
L.~Freidel, D.~Pranzetti and A.-M. Raclariu, \emph{{Higher spin dynamics in
  gravity and w1+\ensuremath{\infty} celestial symmetries}},
  \href{http://dx.doi.org/10.1103/PhysRevD.106.086013}{\emph{Phys. Rev. D}
  {\bfseries 106} (2022) 086013},
  [\href{https://arxiv.org/abs/2112.15573}{{\ttfamily 2112.15573}}].

\bibitem{Donnay:2019jiz}
L.~Donnay and C.~Marteau, \emph{{Carrollian Physics at the Black Hole
  Horizon}}, \href{http://dx.doi.org/10.1088/1361-6382/ab2fd5}{\emph{Class.
  Quant. Grav.} {\bfseries 36} (2019) 165002},
  [\href{https://arxiv.org/abs/1903.09654}{{\ttfamily 1903.09654}}].

\bibitem{Leblond1965}
J.-M. L\'evy-Leblond, \emph{Une nouvelle limite non-relativiste du groupe de
  {Poincar\'e}}, {\emph{Annales de l'I.H.P. Physique th\'eorique} {\bfseries 3}
  (1965) 1--12}.

\bibitem{Ciambelli:2018xat}
L.~Ciambelli, C.~Marteau, A.~C. Petkou, P.~M. Petropoulos and K.~Siampos,
  \emph{{Covariant Galilean versus Carrollian hydrodynamics from relativistic
  fluids}}, \href{http://dx.doi.org/10.1088/1361-6382/aacf1a}{\emph{Class.
  Quant. Grav.} {\bfseries 35} (2018) 165001},
  [\href{https://arxiv.org/abs/1802.05286}{{\ttfamily 1802.05286}}].

\bibitem{Ciambelli:2018ojf}
L.~Ciambelli and C.~Marteau, \emph{{Carrollian conservation laws and Ricci-flat
  gravity}}, \href{http://dx.doi.org/10.1088/1361-6382/ab0d37}{\emph{Class.
  Quant. Grav.} {\bfseries 36} (2019) 085004},
  [\href{https://arxiv.org/abs/1810.11037}{{\ttfamily 1810.11037}}].

\bibitem{Petkou:2022bmz}
A.~C. Petkou, P.~M. Petropoulos, D.~R. Betancour and K.~Siampos,
  \emph{{Relativistic Fluids, Hydrodynamic Frames and Their Galilean Versus
  Carrollian Avatars}},
  \href{http://dx.doi.org/10.1007/JHEP09(2022)162}{\emph{JHEP} {\bfseries 09}
  (2022) 162}, [\href{https://arxiv.org/abs/2205.09142}{{\ttfamily
  2205.09142}}].

\bibitem{Freidel:2022bai}
L.~Freidel and P.~Jai-akson, \emph{{Carrollian Hydrodynamics from Symmetries}},
   [\href{https://arxiv.org/abs/2209.03328}{{\ttfamily 2209.03328}}].

\bibitem{Penna:2018gfx}
R.~F. Penna, \emph{{Near-Horizon Carroll Symmetry and Black Hole Love
  Numbers}},  [\href{https://arxiv.org/abs/1812.05643}{{\ttfamily
  1812.05643}}].

\bibitem{Brown:1992br}
J.~D. Brown and J.~W. York, Jr., \emph{{Quasilocal Energy and Conserved Charges
  Derived from the Gravitational Action}},
  \href{http://dx.doi.org/10.1103/PhysRevD.47.1407}{\emph{Phys. Rev. D}
  {\bfseries 47} (1993) 1407--1419},
  [\href{https://arxiv.org/abs/gr-qc/9209012}{{\ttfamily gr-qc/9209012}}].

\bibitem{Brown:2000dz}
J.~D. Brown, S.~R. Lau and J.~W. York, Jr., \emph{{Action and Energy of the
  Gravitational Field}},
  [\href{https://arxiv.org/abs/gr-qc/0010024}{{\ttfamily gr-qc/0010024}}].

\bibitem{Bekaert:2015xua}
X.~Bekaert and K.~Morand, \emph{{Connections and dynamical trajectories in
  generalised Newton-Cartan gravity II. An ambient perspective}},
  \href{http://dx.doi.org/10.1063/1.5030328}{\emph{J. Math. Phys.} {\bfseries
  59} (2018) 072503}, [\href{https://arxiv.org/abs/1505.03739}{{\ttfamily
  1505.03739}}].

\bibitem{Figueroa-OFarrill:2019sex}
J.~Figueroa-O'Farrill, R.~Grassie and S.~Prohazka, \emph{{Geometry and BMS Lie
  algebras of spatially isotropic homogeneous spacetimes}},
  \href{http://dx.doi.org/10.1007/JHEP08(2019)119}{\emph{JHEP} {\bfseries 08}
  (2019) 119}, [\href{https://arxiv.org/abs/1905.00034}{{\ttfamily
  1905.00034}}].

\bibitem{Figueroa-OFarrill:2020gpr}
J.~Figueroa-O'Farrill, \emph{{On the intrinsic torsion of spacetime
  structures}},  [\href{https://arxiv.org/abs/2009.01948}{{\ttfamily
  2009.01948}}].

\bibitem{Herfray:2021qmp}
Y.~Herfray, \emph{{Carrollian Manifolds and Null Infinity: a View from Cartan
  Geometry}}, \href{http://dx.doi.org/10.1088/1361-6382/ac635f}{\emph{Class.
  Quant. Grav.} {\bfseries 39} (2022) 215005},
  [\href{https://arxiv.org/abs/2112.09048}{{\ttfamily 2112.09048}}].

\bibitem{Ashtekar:2021wld}
A.~Ashtekar, N.~Khera, M.~Kolanowski and J.~Lewandowski, \emph{{Non-Expanding
  Horizons: Multipoles and the Symmetry Group}},
  \href{http://dx.doi.org/10.1007/JHEP01(2022)028}{\emph{JHEP} {\bfseries 01}
  (2022) 028}, [\href{https://arxiv.org/abs/2111.07873}{{\ttfamily
  2111.07873}}].

\bibitem{Baiguera:2022lsw}
S.~Baiguera, G.~Oling, W.~Sybesma and B.~T. S\o{}gaard, \emph{{Conformal
  Carroll Scalars with Boosts}},
  [\href{https://arxiv.org/abs/2207.03468}{{\ttfamily 2207.03468}}].

\bibitem{Hansen:2021fxi}
D.~Hansen, N.~A. Obers, G.~Oling and B.~T. S\o{}gaard, \emph{{Carroll Expansion
  of General Relativity}},  [\href{https://arxiv.org/abs/2112.12684}{{\ttfamily
  2112.12684}}].

\bibitem{Donnay:2015abr}
L.~Donnay, G.~Giribet, H.~A. Gonzalez and M.~Pino, \emph{{Supertranslations and
  Superrotations at the Black Hole Horizon}},
  \href{http://dx.doi.org/10.1103/PhysRevLett.116.091101}{\emph{Phys. Rev.
  Lett.} {\bfseries 116} (2016) 091101},
  [\href{https://arxiv.org/abs/1511.08687}{{\ttfamily 1511.08687}}].

\bibitem{Donnay:2016ejv}
L.~Donnay, G.~Giribet, H.~A. Gonz\'alez and M.~Pino, \emph{{Extended Symmetries
  at the Black Hole Horizon}},
  \href{http://dx.doi.org/10.1007/JHEP09(2016)100}{\emph{JHEP} {\bfseries 09}
  (2016) 100}, [\href{https://arxiv.org/abs/1607.05703}{{\ttfamily
  1607.05703}}].

\bibitem{Chandrasekaran:2018aop}
V.~Chandrasekaran, E.~E. Flanagan and K.~Prabhu, \emph{{Symmetries and charges
  of general relativity at null boundaries}},
  \href{http://dx.doi.org/10.1007/JHEP11(2018)125}{\emph{JHEP} {\bfseries 11}
  (2018) 125}, [\href{https://arxiv.org/abs/1807.11499}{{\ttfamily
  1807.11499}}].

\bibitem{Hopfmuller:2018fni}
F.~Hopfm{\"u}ller and L.~Freidel, \emph{{Null Conservation Laws for Gravity}},
  \href{http://dx.doi.org/10.1103/PhysRevD.97.124029}{\emph{Phys. Rev. D}
  {\bfseries 97} (2018) 124029},
  [\href{https://arxiv.org/abs/1802.06135}{{\ttfamily 1802.06135}}].

\bibitem{Jafari:2019bpw}
G.~Jafari, \emph{{Stress Tensor on Null Boundaries}},
  \href{http://dx.doi.org/10.1103/PhysRevD.99.104035}{\emph{Phys. Rev. D}
  {\bfseries 99} (2019) 104035},
  [\href{https://arxiv.org/abs/1901.04054}{{\ttfamily 1901.04054}}].

\bibitem{Chandrasekaran:2020wwn}
V.~Chandrasekaran and A.~J. Speranza, \emph{{Anomalies in gravitational charge
  algebras of null boundaries and black hole entropy}},
  \href{http://dx.doi.org/10.1007/JHEP01(2021)137}{\emph{JHEP} {\bfseries 01}
  (2021) 137}, [\href{https://arxiv.org/abs/2009.10739}{{\ttfamily
  2009.10739}}].

\bibitem{Adami:2021nnf}
H.~Adami, D.~Grumiller, M.~M. Sheikh-Jabbari, V.~Taghiloo, H.~Yavartanoo and
  C.~Zwikel, \emph{{Null boundary phase space: slicings, news \& memory}},
  \href{http://dx.doi.org/10.1007/JHEP11(2021)155}{\emph{JHEP} {\bfseries 11}
  (2021) 155}, [\href{https://arxiv.org/abs/2110.04218}{{\ttfamily
  2110.04218}}].

\bibitem{Adami:2021kvx}
H.~Adami, M.~M. Sheikh-Jabbari, V.~Taghiloo and H.~Yavartanoo, \emph{{Null
  Surface Thermodynamics}},
  \href{http://dx.doi.org/10.1103/PhysRevD.105.066004}{\emph{Phys. Rev. D}
  {\bfseries 105} (2022) 066004},
  [\href{https://arxiv.org/abs/2110.04224}{{\ttfamily 2110.04224}}].

\bibitem{Chandrasekaran:2021hxc}
V.~Chandrasekaran, E.~E. Flanagan, I.~Shehzad and A.~J. Speranza,
  \emph{{Brown-York Charges at Null Boundaries}},
  \href{http://dx.doi.org/10.1007/JHEP01(2022)029}{\emph{JHEP} {\bfseries 01}
  (2022) 029}, [\href{https://arxiv.org/abs/2109.11567}{{\ttfamily
  2109.11567}}].

\bibitem{Ciambelli:2019lap}
L.~Ciambelli, R.~G. Leigh, C.~Marteau and P.~M. Petropoulos, \emph{{Carroll
  Structures, Null Geometry and Conformal Isometries}},
  \href{http://dx.doi.org/10.1103/PhysRevD.100.046010}{\emph{Phys. Rev. D}
  {\bfseries 100} (2019) 046010},
  [\href{https://arxiv.org/abs/1905.02221}{{\ttfamily 1905.02221}}].

\bibitem{Duval:2014uva}
C.~Duval, G.~W. Gibbons and P.~A. Horv\'athy, \emph{{Conformal Carroll Groups
  and Bms Symmetry}},
  \href{http://dx.doi.org/10.1088/0264-9381/31/9/092001}{\emph{Class. Quant.
  Grav.} {\bfseries 31} (2014) 092001},
  [\href{https://arxiv.org/abs/1402.5894}{{\ttfamily 1402.5894}}].

\bibitem{Duval:2014lpa}
C.~Duval, G.~W. Gibbons and P.~A. Horv\'athy, \emph{{Conformal Carroll
  Groups}}, \href{http://dx.doi.org/10.1088/1751-8113/47/33/335204}{\emph{J.
  Phys. A} {\bfseries 47} (2014) 335204},
  [\href{https://arxiv.org/abs/1403.4213}{{\ttfamily 1403.4213}}].

\bibitem{Bagchi:2019xfx}
A.~Bagchi, A.~Mehra and P.~Nandi, \emph{{Field Theories with Conformal
  Carrollian Symmetry}},
  \href{http://dx.doi.org/10.1007/JHEP05(2019)108}{\emph{JHEP} {\bfseries 05}
  (2019) 108}, [\href{https://arxiv.org/abs/1901.10147}{{\ttfamily
  1901.10147}}].

\bibitem{Bagchi:2022emh}
A.~Bagchi, S.~Banerjee, R.~Basu and S.~Dutta, \emph{{Scattering Amplitudes:
  Celestial and Carrollian}},
  \href{http://dx.doi.org/10.1103/PhysRevLett.128.241601}{\emph{Phys. Rev.
  Lett.} {\bfseries 128} (2022) 241601},
  [\href{https://arxiv.org/abs/2202.08438}{{\ttfamily 2202.08438}}].

\bibitem{Campoleoni:2022wmf}
A.~Campoleoni, L.~Ciambelli, A.~Delfante, C.~Marteau, P.~M. Petropoulos and
  R.~Ruzziconi, \emph{{Holographic Lorentz and Carroll Frames}},
  [\href{https://arxiv.org/abs/2208.07575}{{\ttfamily 2208.07575}}].

\bibitem{Bagchi:2022owq}
A.~Bagchi, D.~Grumiller and P.~Nandi, \emph{{Carrollian Superconformal Theories
  and Super Bms}}, \href{http://dx.doi.org/10.1007/JHEP05(2022)044}{\emph{JHEP}
  {\bfseries 05} (2022) 044},
  [\href{https://arxiv.org/abs/2202.01172}{{\ttfamily 2202.01172}}].

\bibitem{Donnay:2022aba}
L.~Donnay, A.~Fiorucci, Y.~Herfray and R.~Ruzziconi, \emph{{Carrollian
  Perspective on Celestial Holography}},
  \href{http://dx.doi.org/10.1103/PhysRevLett.129.071602}{\emph{Phys. Rev.
  Lett.} {\bfseries 129} (2022) 071602},
  [\href{https://arxiv.org/abs/2202.04702}{{\ttfamily 2202.04702}}].

\bibitem{Donnay:2022sdg}
L.~Donnay, S.~Pasterski and A.~Puhm, \emph{{Goldilocks Modes and the Three
  Scattering Bases}},
  \href{http://dx.doi.org/10.1007/JHEP06(2022)124}{\emph{JHEP} {\bfseries 06}
  (2022) 124}, [\href{https://arxiv.org/abs/2202.11127}{{\ttfamily
  2202.11127}}].

\bibitem{Gupta:2020dtl}
N.~Gupta and N.~V. Suryanarayana, \emph{{Constructing Carrollian CFTs}},
  \href{http://dx.doi.org/10.1007/JHEP03(2021)194}{\emph{JHEP} {\bfseries 03}
  (2021) 194}, [\href{https://arxiv.org/abs/2001.03056}{{\ttfamily
  2001.03056}}].

\bibitem{Bagchi:2022eav}
A.~Bagchi, A.~Banerjee, S.~Dutta, K.~S. Kolekar and P.~Sharma, \emph{{Carroll
  Covariant Scalar Fields in Two Dimensions}},
  [\href{https://arxiv.org/abs/2203.13197}{{\ttfamily 2203.13197}}].

\bibitem{Mars:1993mj}
M.~Mars and J.~M.~M. Senovilla, \emph{{Geometry of General Hypersurfaces in
  Space-Time: Junction Conditions}},
  \href{http://dx.doi.org/10.1088/0264-9381/10/9/026}{\emph{Class. Quant.
  Grav.} {\bfseries 10} (1993) 1865--1897},
  [\href{https://arxiv.org/abs/gr-qc/0201054}{{\ttfamily gr-qc/0201054}}].

\bibitem{Mars:2013qaa}
M.~Mars, \emph{{Constraint equations for general hypersurfaces and applications
  to shells}}, \href{http://dx.doi.org/10.1007/s10714-013-1579-9}{\emph{Gen.
  Rel. Grav.} {\bfseries 45} (2013) 2175--2221},
  [\href{https://arxiv.org/abs/1303.4575}{{\ttfamily 1303.4575}}].

\bibitem{Parikh:1997ma}
M.~Parikh and F.~Wilczek, \emph{{An Action for Black Hole Membranes}},
  \href{http://dx.doi.org/10.1103/PhysRevD.58.064011}{\emph{Phys. Rev. D}
  {\bfseries 58} (1998) 064011},
  [\href{https://arxiv.org/abs/gr-qc/9712077}{{\ttfamily gr-qc/9712077}}].

\bibitem{Parattu:2015gga}
K.~Parattu, S.~Chakraborty, B.~R. Majhi and T.~Padmanabhan, \emph{{A Boundary
  Term for the Gravitational Action with Null Boundaries}},
  \href{http://dx.doi.org/10.1007/s10714-016-2093-7}{\emph{Gen. Rel. Grav.}
  {\bfseries 48} (2016) 94},
  [\href{https://arxiv.org/abs/1501.01053}{{\ttfamily 1501.01053}}].

\bibitem{Parattu:2016trq}
K.~Parattu, S.~Chakraborty and T.~Padmanabhan, \emph{{Variational Principle for
  Gravity with Null and Non-null boundaries: A Unified Boundary Counter-term}},
  \href{http://dx.doi.org/10.1140/epjc/s10052-016-3979-y}{\emph{Eur. Phys. J.
  C} {\bfseries 76} (2016) 129},
  [\href{https://arxiv.org/abs/1602.07546}{{\ttfamily 1602.07546}}].

\bibitem{Hopfmuller:2016scf}
F.~Hopfm{\"u}ller and L.~Freidel, \emph{{Gravity Degrees of Freedom on a Null
  Surface}}, \href{http://dx.doi.org/10.1103/PhysRevD.95.104006}{\emph{Phys.
  Rev. D} {\bfseries 95} (2017) 104006},
  [\href{https://arxiv.org/abs/1611.03096}{{\ttfamily 1611.03096}}].

\bibitem{Gourgoulhon:2007ue}
E.~Gourgoulhon, \emph{{3+1 Formalism and Bases of Numerical Relativity}},
  [\href{https://arxiv.org/abs/gr-qc/0703035}{{\ttfamily gr-qc/0703035}}].

\bibitem{Gourgoulhon:2005ng}
E.~Gourgoulhon and J.~L. Jaramillo, \emph{{A 3+1 Perspective on Null
  Hypersurfaces and Isolated Horizons}},
  \href{http://dx.doi.org/10.1016/j.physrep.2005.10.005}{\emph{Phys. Rept.}
  {\bfseries 423} (2006) 159--294},
  [\href{https://arxiv.org/abs/gr-qc/0503113}{{\ttfamily gr-qc/0503113}}].

\bibitem{Brady:1995na}
P.~R. Brady, S.~Droz, W.~Israel and S.~M. Morsink, \emph{{Covariant Double Null
  Dynamics: (2+2) Splitting of the Einstein Equations}},
  \href{http://dx.doi.org/10.1088/0264-9381/13/8/015}{\emph{Class. Quant.
  Grav.} {\bfseries 13} (1996) 2211--2230},
  [\href{https://arxiv.org/abs/gr-qc/9510040}{{\ttfamily gr-qc/9510040}}].

\bibitem{Bondi:1962px}
H.~Bondi, M.~G.~J. van~der Burg and A.~W.~K. Metzner, \emph{{Gravitational
  waves in general relativity. 7. Waves from axisymmetric isolated systems}},
  \href{http://dx.doi.org/10.1098/rspa.1962.0161}{\emph{Proc. Roy. Soc. Lond.}
  {\bfseries A269} (1962) 21--52}.

\bibitem{Sachs:1962wk}
R.~K. Sachs, \emph{{Gravitational waves in general relativity. 8. Waves in
  asymptotically flat space-times}},
  \href{http://dx.doi.org/10.1098/rspa.1962.0206}{\emph{Proc. Roy. Soc. Lond.}
  {\bfseries A270} (1962) 103--126}.

\bibitem{Freidel:2021fxf}
L.~Freidel, R.~Oliveri, D.~Pranzetti and S.~Speziale, \emph{{The Weyl BMS group
  and Einstein\textquoteright{}s equations}},
  \href{http://dx.doi.org/10.1007/JHEP07(2021)170}{\emph{JHEP} {\bfseries 07}
  (2021) 170}, [\href{https://arxiv.org/abs/2104.05793}{{\ttfamily
  2104.05793}}].

\bibitem{Newman:1962cia}
E.~T. Newman and T.~W.~J. Unti, \emph{{Behavior of Asymptotically Flat Empty
  Spaces}}, \href{http://dx.doi.org/10.1063/1.1724303}{\emph{J. Math. Phys.}
  {\bfseries 3} (1962) 891}.

\bibitem{Duval:2014uoa}
C.~Duval, G.~W. Gibbons, P.~A. Horvathy and P.~M. Zhang, \emph{{Carroll versus
  Newton and Galilei: two dual non-Einsteinian concepts of time}},
  \href{http://dx.doi.org/10.1088/0264-9381/31/8/085016}{\emph{Class. Quant.
  Grav.} {\bfseries 31} (2014) 085016},
  [\href{https://arxiv.org/abs/1402.0657}{{\ttfamily 1402.0657}}].

\bibitem{Duggal}
K.~L. Duggal and A.~Bejancu, \emph{Lightlike Submanifolds of Semi-Riemannian
  Manifolds and Applications}.
\newblock Kluwer Academic,, Dordrecht, 1996.

\bibitem{Gibbons:2008zi}
G.~W. Gibbons, C.~A.~R. Herdeiro, C.~M. Warnick and M.~C. Werner,
  \emph{{Stationary Metrics and Optical Zermelo-Randers-Finsler Geometry}},
  \href{http://dx.doi.org/10.1103/PhysRevD.79.044022}{\emph{Phys. Rev. D}
  {\bfseries 79} (2009) 044022},
  [\href{https://arxiv.org/abs/0811.2877}{{\ttfamily 0811.2877}}].

\bibitem{forthcoming}
L.~Freidel and P.~Jai-akson, \emph{{Symmetries and Einstein equations on null
  boundaries}},  [\href{https://arxiv.org/abs/in preparation}{{\ttfamily in
  preparation}}].

\bibitem{Freidel:2021dxw}
L.~Freidel, \emph{{A Canonical Bracket for Open Gravitational System}},
  [\href{https://arxiv.org/abs/2111.14747}{{\ttfamily 2111.14747}}].

\end{thebibliography}\endgroup
\bibliographystyle{Biblio}

\end{document}